\documentclass[twocolumn]{aastex62}
\usepackage{graphics,epsf}
\usepackage{amsmath}                
\usepackage{amsfonts}               
\usepackage{amssymb}                
\usepackage{epsfig}                 
\usepackage{graphicx}
\usepackage{float}
\usepackage{color}
\usepackage[para,online,flushleft]{threeparttable}

\newcommand{\cm}{{~\rm cm}}
\newcommand{\km}{{~\rm km}}
\newcommand{\s}{{~\rm s}}

\newcommand{\g}{{~\rm g}}

\newcommand{\K}{{~\rm K}}
\newcommand{\erg}{{~\rm erg}}
\newcommand{\yr}{{~\rm yr}}

	\def \aap{A\&A}
	\def \araa{ARA\&A}
	\def \apj{ApJ}
	\def \apjl{ApJ}
	\def \apjs{ApJS}

	\def \mnras{MNRAS}

\begin{document}

\title{Simulating highly-eccentric common envelope jets supernova (CEJSN) impostors}

\author{Ron Schreier}
\affiliation{Department of Physics, Technion, Haifa, 3200003, Israel; \\ ronsr@technion.ac.il, shlomi.hillel@gmail.com, soker@physics.technion.ac.il}

\author{Shlomi Hillel}
\affiliation{Department of Physics, Technion, Haifa, 3200003, Israel; \\ ronsr@technion.ac.il, shlomi.hillel@gmail.com, soker@physics.technion.ac.il}

\author[0000-0001-6107-0887]{Sagiv Shiber}
\affiliation{Department of Physics and Astronomy, Louisiana State University, Baton Rouge, LA, 70803 USA; sshiber1@lsu.edu}

\author[0000-0003-0375-8987]{Noam Soker}
\affiliation{Department of Physics, Technion, Haifa, 3200003, Israel; \\ ronsr@technion.ac.il, shlomi.hillel@gmail.com, soker@physics.technion.ac.il}
\affiliation{Guangdong Technion Israel Institute of Technology, Shantou 515069, Guangdong Province, China}

\begin{abstract}
We conduct three-dimensional hydrodynamical simulations of eccentric common envelope jets supernova (CEJSN) impostors, i.e., a neutron star (NS) that crosses through the envelope of a red supergiant star on a highly eccentric orbit and launches jets as it accretes mass from the envelope. Because of numerical limitations we apply a simple prescription where we inject the assumed jets' power into two opposite conical regions inside the envelope. We find the outflow morphology to be very complicated, clumpy, and non-spherical, having a large-scale symmetry only about the equatorial plane. The outflow morphology can substantially differ between simulations that differ by their jets' power. We estimate by simple means the light curve to be very bumpy, to have a rise time of one to a few months, and to slowly decay in about a year to several years. These eccentric CEJSN impostors will be classified as `gap' objects, i.e., having a luminosity between those of classical novae and typical supernovae (termed also ILOTs for intermediate luminosity optical transients). We strengthen a previous conclusion that CEJSN impostors might account for some peculiar ILOTs, in particular those that might repeat over timescales of months to years. 
\end{abstract}


\section{INTRODUCTION}
\label{sec:intro}

A common envelope jet supernova (CEJSN) event is a jet-driven  energetic transient that mimic in many respects a core collapse supernova (CCSN). The basic CEJSN scenario is of a neutron star (NS) or a black hole (BH) that enter the envelope of a giant star, accrete mass through an accretion disk that launches jets, and inspiral in a common envelope evolution (CEE). We will concentrate here on a NS companion, but most processes hold also for a BH companion. If the NS enters the core that is much denser than the giant envelope, it accretes at a much higher rate and launches very energetic jets that explode the star. This is a CEJSN event (e.g., \citealt{Gilkisetal2019, Sokeretal2019CEJSN, GrichenerSoker2019a, Schroderetal2020, GrichenerSoker2021, Soker2021Triple}). In the final phase the NS tidally destroys the core to form a massive accretion disk around the NS.  \cite{Papishetal2015} claim that this process prevents the formation of Thorne-\.{Z}ytkow objects. 

The rapid neutrino cooling of the accreted gas at high accretion rates of  $\dot M_{\rm acc} \ga 10^{-3} M_\odot \yr^{-1}$ \citep{HouckChevalier1991, Chevalier1993, Chevalier2012} and the formation of an accretion disk around the very compact NS or BH (e.g.,  \citealt{ArmitageLivio2000, Papishetal2015, SokerGilkis2018}) allow the CEJSN process. 

Past studies disagree on the exact mass accretion rate relative to the Bondi-Hoyle-Lyttleton (BHL) accretion rate $\dot M_{\rm BHL}$ and on whether the accretion process is through an accretion disk or not (e.g., \citealt{RasioShapiro1991, Fryeretal1996, Lombardietal2006, RickerTaam2008, Shiberetal2016, MacLeodRamirezRuiz2015a, MacLeodRamirezRuiz2015b,  MacLeodetal2017}). We adopt recent studies that show the NS mass accretion rate to be $\dot M_{\rm acc} \approx 0.1 - 0.2 \dot M_{\rm BHL}$ \citep{LopezCamaraetal2019, LopezCamaraetal2020MN}, and recent studies that support the formation of an accretion disk around the companion in CEE (e.g., \citealt{Chamandyetal2018}), in particular around a NS (\citealt{LopezCamaraetal2020MN}). We will also take into account the negative jet feedback mechanism in CEE (e.g., \citealt{Soker2016Rev}), wherein jets remove mass from the surrounding of the mass-accreting companion, and consider much lower accretion rates as well. Overall, neutrinos carry most of the energy that the accretion process liberates, and jets carry the rest. The equation of state of the NS also influences the outcome of accretion (e.g., \citealt{Holgadoetal2021}). 

The variety of CEJSN types as determined by the giant envelope mass and its spin during the CEE, by the presence of a NS or a BH companion, and by whether the jets at the final phase break out from the envelope or not, lead to different types of light curves \citep{Sokeretal2019CEJSN} that might account for some puzzling supernovae (SNe). These SNe include the enigmatic SN~iPTF14hls and SN~2020faa \citep{SokerGilkis2018}, the fast-rising blue optical transient AT2018cow \citep{Sokeretal2019CEJSN}, and the SNe  
SN1979c and SN1998s \citep{Schroderetal2020}.
   
As well, the jets in some CEJSNe with a NS accreting mass from a CO core of a red supergiant (RSG) might be an r-process nucleosynthesis site \citep{Papishetal2015, GrichenerSoker2019a, GrichenerSoker2019b}, while a BH accreting from the envelope of a RSG might be one of the sources of high-energy neutrinos \citep{GrichenerSoker2020}. 
The efficient envelope mass removal by jets in a CEE (e.g., \citealt{Shiberetal2019}) can increase the CEE efficiency parameter above unity, i.e., $\alpha_{\rm CE} > 1$, as some scenarios require (e.g. \citealt{Fragosetal2019, Zevinetal2021, Garciaetal2021}).

To this rich variety of CEJSN types when we consider binary systems, studies added recently CEJSN cases that result from triple star systems (e.g., \citealt{Soker2021Triple, Soker2021NSNS}). One of these triple-star cases is the eccentric CEJSN impostor that is the subject of the present study. In a CEJSN impostor \citep{Gilkisetal2019} the NS (or BH) accretes mass from the tenuous envelope to power a bright transient event. However, it does not spiral-in all the way to the core. One setting to form an eccentric CEJSN impostor is by a tertiary star on a wide orbit that perturbed the NS-RSG inner binary. 
The tertiary star serves only to perturb the NS orbit such that it acquires a highly eccentric orbit and enters the RSG envelope and exits from it \citep{Gilkisetal2019}. While inside the envelope it accretes mass via an accretion disk and launches jets that power a bright event, an eccentric  CEJSN impostor. The process might repeat itself, and the system might later enter a continues CEE toward a CEJSN (if the NS enters the core of the RSG). 

Our goal is to explore some properties of the outflow that the jets induce in an eccentric CEJSN impostor. We describe the numerical scheme in section \ref{sec:Numerical}, and the properties of the jet-driven outflow in section \ref{sec:Results}.  We summarise in section \ref{sec:Summary}.

\section{Numerical set up}
\label{sec:Numerical}

\subsection{Basic simplifying assumptions}
\label{subsec:assumptions}

In this study a NS orbits a RSG at a highly eccentric ellipse and with an orbital period of $T_{\rm orb}=16.6 \yr$. We assume that as it enters the RSG envelope the NS accretes mass through an accretion disk and launches powerful jets. The jet-launching phase lasts for $t_{\rm jets} = 0.51 \yr$. These jets induce mass-loss from the envelope that continues long after the NS has left the RSG envelope. To be able to follow the outflow at later times we make some basic assumptions as follows.

\subsubsection{1-D stellar model}
\label{subsubsec:spherical}

We take the three dimensional (3D) RSG model from the 1D stellar evolution code \texttt{MESA} (\citealt{Paxtonetal2011, Paxtonetal2013, Paxtonetal2015, Paxtonetal2018, Paxtonetal2019}). We evolved a star of zero age main sequence mass of $M_{\rm 1,ZAMS}=15 M_\odot$
for $1.1\times10^6 \yr$, as it reaches a radius of $R_{RSG}=881\,R_{\odot}$ in its RSG phase. Its effective temperature then is $T_{\rm eff}= 3160K$, and due to mass loss its mass at this time is $M_1=  12.5 M_\odot$. 

We place the non-rotating RSG model at the center of the 3D numerical grid of the hydrodynamical code {\sc flash} \citep{Fryxelletal2000}. 
The outer boundary of the initial RSG model has a density of  $\rho(R_{RSG}) = 2.1 \times 10^{-9} \g \cm^{-3}$. To prevent numerical instabilities and too short time steps we set the initial density of the numerical grid outside the RSG model to have a density of $\rho_{\rm grid,0} = 2.1 \times 10^{-13} \g \cm^{-3}$ and a temperature of $T_{\rm grid,0}= 1000 \K$. 
This compact circumstellar mater is not unrealistic, as the total mass in this zone is few$\times 0.1 M_\odot$ and it might represent an effervescent zone \citep{Soker2021effervescent}. 
The 1D stellar model has a metallicity of $Z=0.02$, but in the 3D hydrodynamical simulation we use a pure hydrogen and assume it to be fully ionised. 

During the simulations that last for no more than several years, the inner volume of the star does not evolve much and it is not affected much by the jets. To prevent too short time steps we take the volume inner to the radius $R_{\rm in} = 0.2R_{RSG} = 176\,R_{\odot}$ and of mass $M_{\rm in} = 5.65 M_{\odot}$ to be an inert sphere having constant density, pressure, and temperature. We do include the gravitational potential of this inner inert sphere.
The 1D stellar model has a convective envelope, but in placing the 1D stellar model in the 3D numerical grid we do not include any convection motion. 

For the duration of our simulations, which is about few dynamical times, regions of the stellar model do not change much until the shocks that the jets excite hit them.
In a test simulation without jets there is a very weak inflow into the grid. Within a time scale of $10^8 \s$, which is the duration of our simulations (beside extending the time to $2 \times 10^8 \s$ to reveal the light curve), the mass in the grid increases by $4 \%$. Since the main effect of the jets is mass removal, this has no influence on our results to the accuracy we demand from the simulations. In the test simulation without jets small-scale perturbations on the stellar surface appear at about $ 10^7 \s$, and reach a typical size of $\simeq 5\%$ of the stellar radius at $t=2 \times 10^7 \s$.  The envelope of the RSG is convectively-unstable and so we expect the envelope in the numerical grid to be unstable. However, in most of the envelope the profile is only slightly above adiabatic. Therefore, convection takes time to develop large amplitudes, as we indeed find. Only in the outer envelope zones that contain little mass does the profile becomes super-adiabatic and instabilities develop somewhat faster.  As we show later, by the time of $2 \times 10^7 \s$ the shocks have already transverse the entire star.  Only at $1.5 \times 10^8 \s$ in the simulation without jets, which is about ten times the shock-transverse time through the envelope, do large scale perturbations that change the large scale envelope structure into a structure that has the imprints of the Cartesian grid appear. 

We conclude that despite the convectively-unstable envelope, the replacement of the 1D stellar model solar composition with pure hydrogen composition, and the Cartesian numerical grid with its finite resolution, our 3D numerical stellar model maintains its structure to the desired accuracy for the length of our simulations.

\subsubsection{A highly-eccentric orbit}
\label{subsubsec:orbit}

We let the NS orbit inside the RSG as if the RSG is a point mass. The periastron and apastron of this elliptical orbit are at $r_{\rm p}=400 R_\odot$ and $r_{\rm a}=6400 R_\odot$, respectively, i.e., an eccentricity of $e=0.8824$. Because we assume that the NS enters the envelope for the first time as a result of a perturbation by a tertiary star, pre-CEE effects on the eccentricity (e.g., \citealt{Vicketal2021MNRAS}) and effects of continues eccentric CEE (e.g.,  \citealt{GlanzPerets2021}) are not significant.

For the hydrodynamical simulations we do take a spherical gravitational field of $g_r(r)=-GM_0(r)/r^2$, where $M_0(r)$ is the mass of the RSG inner to radius $r$ at $t=0$. Namely, $g_r(r)$ does not change with time.
This implies that at $t=0$ the star in our 3D numerical grid is in hydrostatic equilibrium to a large accuracy, but not fully so, e.g., because of the Cartesian grid. 
We launch the jets only when the NS is inside the envelope. We neglect the gravity of the NS altogether, as we also neglect the self-gravity of the envelope.  We expect that a more accurate orbit that includes all effects, will not change much our results. The highest uncertainties are in the properties of the jets (mass outflow rate and opening angle; the jets initial velocity is about the escape speed from the NS). 

\subsubsection{The jets' power}
\label{subsubsec:jets}

Because of numerical limitation we do not resolve the close vicinity of the NS as it passes through the envelope, and so we do not follow the accretion process. Resolving the vicinity of the NS would constrain the numerical time step to be very short and would allow us to follow only a short evolutionary time  (e.g., \citealt{MorenoMendezetal2017, LopezCamaraetal2019, LopezCamaraetal2020MN}). 
For the same reason we cannot follow the launching process of the jets by the accretion disk. 
   
We launch the jets manually in two opposite cones that are perpendicular to the equatorial plane and within which the jets' material mixes with the ambient envelope gas (section \ref{subsubsec:interaction}). 
We set the power of the two jets to be a fraction $\zeta$ of the accretion power according to the BHL mass accretion rate $\dot M_{\rm BHL}$
\begin{equation}
\dot E_{\rm 2j} = \zeta \frac {G M_{\rm NS} \dot M_{\rm BHL}}{R_{\rm NS}} \equiv \zeta \dot E_{\rm acc,p,BHL}, 
\label{eq:DotE2j}
\end{equation}
where $M_{\rm NS}=1.4 M_\odot$ and $R_{\rm NS}=12 \km$ are the mass and radius of the NS, respectively, and the second equality defines $\dot E_{\rm acc,p,BHL}$. 
 
We use the BHL mass accretion rate in its simple form for high Mach number flow. In this case the accretion rate goes as $\dot M_{\rm BHL} \propto \rho(r) v^{-3}_{\rm rel}$, where $v_{\rm rel}$ is the relative NS-envelope velocity which we take to be the orbital velocity of the NS. Overall, the power of the two jets together in our specific setting is  
\begin{eqnarray}
\begin{aligned}
\dot E_{\rm 2j} & = 1.25  \times 10^{42} \left( \frac{\zeta}{4.77 \times 10^{-4}} \right)
\left[ \frac{v(r)}{v_{\rm p}} \right]^{-3}
\left[ \frac{\rho(r)}{\rho_{\rm p}} \right]
\\ & \times 
\left( \frac{\dot E_{\rm acc,p,BHL}}{2.62 \times 10^{45} \erg \s^{-1}} \right)
\erg \s^{-1}, 
\label{eq:DotE2jValue}
\end{aligned}
\end{eqnarray}
where $v_{\rm p}$ and $\rho_{\rm p}$ are the velocity of the NS and the envelope density at periastron. 

A typical fraction of the accreted mass that the jets carry out (at about the escape velocity) is $\approx 0.1$. However, because the jets expel mass from the NS vicinity the jets-envelope interaction operates in a negative feedback mechanism (e.g.,  \citealt{Soker2016Rev} for a review). This implies that for the very deep potential well of a NS the jets are efficient in removing mass and so the mass accretion rate is much lower than the BHL mass accretion rate, $\dot M_{\rm acc} \ll \dot M_{\rm BHL}$ (e.g., \citealt{Gricheneretal2021}). This is the reason we take in this study $\zeta \ll 0.1$.   

\begin{figure} 
\centering
\includegraphics[width=0.42\textwidth]{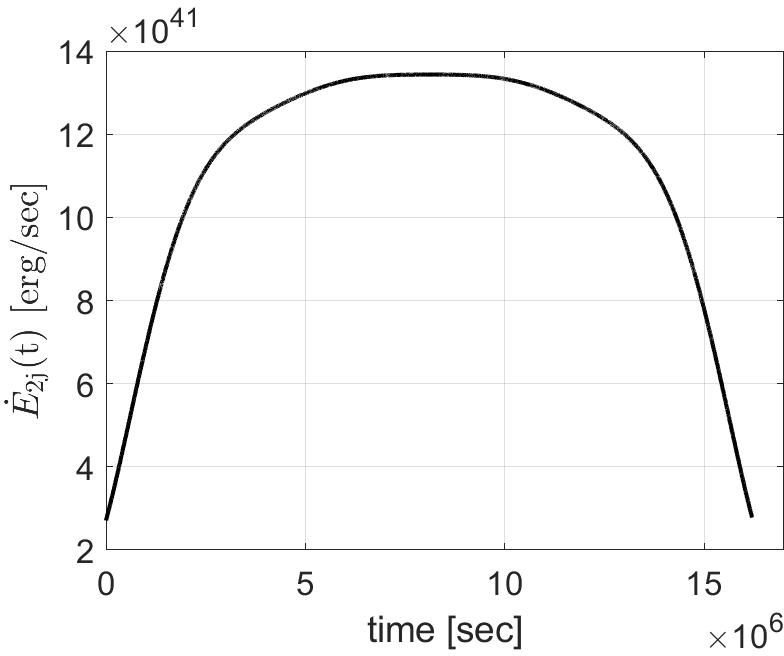}
\caption{The power of the two jets according to equation (\ref{eq:DotE2jValue}) and for $\zeta=4.77 \times 10^{-4}$ (simulation J42/G2) as function of time from when the NS enters the RSG envelope until it gets out. The total energy the jets carry is $1.82 \times 10^{49} \erg$ for this case. Energy and power are linear with $\zeta$. 
}  
\label{fig:jetspower}
\end{figure}

In Fig. \ref{fig:jetspower} we present the power of the two jets together as function of time during the time the NS is inside the RSG envelope, where at t=0 the NS enters the RSG envelope. We assume the jets' power according to density of the unperturbed envelope, i.e., the value of $\rho(r)$ in equation (\ref{eq:DotE2jValue}) is that of the initial stellar model. The global effect of the jets' in reducing the accretion rate is in the value of $\zeta \ll 1$.   

\subsubsection{The jet-envelope interaction in the NS vicinity}
\label{subsubsec:interaction}

To overcome numerical limitations (we cannot have too short time-steps and are unable to have the required resolution near the NS) we use subgrid calculation for the interaction of the jets with the envelope in the NS vicinity. 
We take each of the two opposite jets to inflate a conically-shaped lobe with a length of $L_{\rm L} = 7\times 10^{12} \cm$ and a half opening angle of $\theta_{\rm L} =30^\circ$.  The length of each lobe and the diameter at the top of the lobe are resolved by about 4 grid-cells.   At the beginning of each numerical time step the total mass, total thermal energy, and total kinetic energy within the volume $V_{\rm 2j}$ of the two lobes are $M_{\rm 2c,0}$, $E_{\rm 2tc,0}$, and $E_{\rm 2kc,0}$, respectively.  
 
Because by far the largest uncertainties are in the properties of the jets, 
to the accuracy of this study we can neglect the mass that the NS accretes from the envelope in its vicinity, 
the mass that the jets add to the envelope in the NS vicinity, 
and the change in the gravitational energy of the envelope gas in the NS vicinity due to the NS gravity. 
Most important to notice is that because the jets velocity is much larger than the sound speed in the envelope,  $v_{\rm j} \gg c_{\rm s,env}$, the main role of the jets is the deposition of energy to the envelope rather than the deposition of momentum. 
Therefore, we consider the jet-envelope interaction to be similar to an explosion and take the kinetic energy and thermal energy in the lobes to be about equal to each other. The outward momentum in the lobes, i.e., a radial velocity $v_{\rm c,r}$ times the mass in the lobes, comes from the pressure gradient that the jets build as they are shocked.

Because of our assumption of an explosion-like interaction, the sound speed in the lobe is about equal to the radial outflow speed in the lobe, $v_{\rm jL}\simeq c_{\rm sL}$. The lobe expands to the side at a speed of $\simeq c_{\rm sL}$ and in the radial direction at $\simeq v_{\rm jL} + c_{\rm sL} \simeq 2 c_{\rm sL} $, so the half opening angle of the lobe obeys $\sin \theta_{\rm L} \simeq 0.5$. This is the reason we take $\theta_{\rm L} =30^\circ$. The half opening angle might be somewhat larger or smaller, but we do not expect narrow lobes or very wide lobes.

Based on these assumptions we write the equations for the properties of the gas in the conical lobes
\begin{eqnarray}
\begin{aligned}
& M_{\rm 2c}=M_{\rm 2c,0}  
\\ &
E_{\rm 2c}=E_{\rm 2kc,0}  + E_{\rm 2tc,0} + \dot E_{\rm 2j} \Delta t
\\ &
E_{\rm 2kc}=0.5E_{\rm 2c}; \qquad E_{\rm 2tc}=0.5E_{\rm 2c};
\\ &
v_{\rm c,r} = \sqrt{2 E_{\rm 2kc}/ M_{\rm 2c} }, 
\end{aligned}
\label{eq:Cones}
\end{eqnarray}
$$$$
for the total mass, the total energy, the total kinetic energy, the total thermal energy, and the radial velocity at the end of each time step  $\Delta t$, 
respectively.
The quantities at the beginning of the time step are marked by subscript `0'. Subscript `2' stands for the two conical outflow zones (lobes) together.  
     
\subsection{The grid}
\label{subsec:grid}
The computational numerical grid is a cube with a side of either (G1) $L_g=5 \times 10^{14} \cm$ with a base grid resolution of $\Delta L_g=L_g/8=6.25\times 10^{13} \cm$, or (G2) $L_g=10^{15} \cm$ with a base grid resolution of $\Delta L_g=L_g/8=1.25\times 10^{14} \cm$.
In both types of grid we set an adaptive-mesh-refinement (AMR) of 6 refinement levels, namely, the smallest grid size is $\Delta_{\rm cell,m}=\Delta L_g/2^5 = L_g/256$.
In the entire computational grid the equation of state of the gas is that of an ideal gas with an adiabatic index of $\gamma=5/3$ plus radiation pressure. The molecular weight throughout the grid is of a pure fully ionised hydrogen $\mu=0.5$. 
In the post-shock zones of the ejecta where density is relatively low 
radiation pressure becomes larger (by a considerable factor) than the thermal pressure of the gas. 
For that, we include in the equation of state both radiation pressure and thermal pressure of the gas,
using the multi-temperature gamma equation of state in the 
{\sc flash} code assuming equal temperature for the radiation, ions, and electrons and an ionised Hydrogen envelope.

The center of the RSG does not change and it is at $(x_1,y_1,z_1)=(0,0,0)$. We take the $z=0$ plane to coincide with the equatorial plane such that periastron is at $(x,y,z,)=(400R_\odot, 0, 0)$. The NS orbits the RSG in the anti-clockwise direction in the figures of the orbital plane that we will present, i.e., the NS enters the RSG envelope from the $y<0$ side and exits from the $y>0$ side.

\subsection{Cases we simulate}
\label{subsec:Cases}

In table \ref{Table:cases} we list the five simulations that we perform. We name them by the power of the two jets combined (equation \ref{eq:DotE2jValue}) and the grid size. 
\begin{table*}[htb!] 
\centering
\begin{tabular}{|c|c|c|c|c|c|c|c|}
\hline
 Case       &$\zeta$              & $\dot E_{\rm 2j,p}$ & $E_{\rm 2j}$ & $M_{\rm out,Ub}$ & $E_{\rm rad}$ & $E_{\rm rad}/E_{\rm 2j}$ & Figures \\ 
            &                     & $[\erg \s^{-1}]$    & $[\erg]$    & $[M_\odot]$      & $[\erg]$      &  &\\   
 \hline 
J39/G1 &  $4.77 \times 10^{-7}$ &$1.25 \times 10^{39}$ & $1.82 \times 10^{46}$ & 0.0014 & $5.5 \times 10^{43}$ & 0.003 & \ref{fig:OutflowRate} \\  
 \hline
J40/G1 &  $4.77 \times 10^{-6}$& $1.25 \times 10^{40}$ & $1.82 \times 10^{47}$ & 0.03 & $1.2 \times 10^{45}$ & 0.007  & \ref{fig:OutflowRate}  \\  
 \hline 
J41/G1 &  $4.77 \times 10^{-5}$& $1.25 \times 10^{41}$ & $1.82 \times 10^{48}$ & 0.58 [0.59] & $2.1 \times 10^{46}$ & 0.016 & \ref{fig:outflow_maps}, \ref{fig:outflow_theta}, \ref{fig:OutflowRate}  \\  
 \hline 
J42/G2&  $4.77 \times 10^{-4}$& $1.25 \times 10^{42}$  & $1.82 \times 10^{49}$ & 2.2 [2.6] & $7.8 \times 10^{47}$ & 0.043 & \ref{fig:jetspower} - \ref{fig:OutflowRate} \\  
 \hline  
J43/G2&  $4.77 \times 10^{-3}$& $1.25 \times 10^{43}$  & $1.82 \times 10^{50}$ & 2.9 [3.2]& $5.2 \times 10^{48}$ & 0.029 & \ref{fig:J42_vs_J43}, \ref{fig:OutflowRate}\\  
 \hline 
  \end{tabular} 
\caption{Summary of the simulations that we perform. The number following the `J' in the name is from the power of 10 in the jets' power at periastron that we give in the third column, and the G1 or G2 in the name stand for the small or large grid, respectively (section \ref{subsec:grid}). In the second column we give the power coefficient $\zeta$ that appears in equations (\ref{eq:DotE2j}) and (\ref{eq:DotE2jValue}). In the third column we list the combined power of the two jets at periastron according to equation (\ref{eq:DotE2jValue}) and in the fourth column the total energy that the two jets deposit to the envelope. We then list the total unbound mass $M_{\rm out,Ub}$ that has left the grid over the time period from $t=0$ to $t=10^8 \s$, and in square parenthesis the amount of unbound mass that has left the grid over the time period from $t=0$ to $2\times 10^8 \s$ in three simulations (see section \ref{subsec:Light Curve}). In the sixth column we list our crude estimates of radiated energy $E_{\rm rad}$ over the time period from $t=0$ to $t=10^8 \s$. 
 }
\label{Table:cases}
\end{table*}

We analyse in more details simulation J42/G2 for which $\zeta=4.77 \times 10^{-4}$. The reason is that we expect this simulation to represent the most realistic case \citep{Gricheneretal2021}. According to the study of \cite{Gricheneretal2021} there are three processes that determine the value of $\zeta$. 
If we ignore the effect of the jets, the accretion rate is somewhat lower than the BHL accretion rate. Based on  different three-dimensional simulations (e.g., \citealt{Livioetal1986, RickerTaam2008, MacLeodRamirezRuiz2015a,  Chamandyetal2018, LopezCamaraetal2020MN}) \cite{Gricheneretal2021} take this reduction factor to be $\xi \approx 0.1- 0.2$.
The jets carry a small fraction $\eta$ of the accretion energy. \cite{Gricheneretal2021} take $\eta \approx 0.1$.
Then there is the effect of the jets on the envelope, i.e., the negative feedback mechanism of the jet-envelope interaction.
\cite{Gricheneretal2021} perform a simple analysis with a spherical model and inject the energy of the jets into the envelope of a RSG model. From that they estimate the reduction factor $\chi_{\rm j}$ in the envelope density due to the effect of the jets, and crudely estimate the effect of the negative jets feedback mechanism to be $\chi_{\rm j} \simeq 0.1-0.2$. 
Overall, \cite{Gricheneretal2021} estimate that 
\begin{equation}
\zeta=\xi \chi_{\rm j} \eta \approx 10^{-3}. 
\label{eq:zeta}
\end{equation} 
This is similar to the value of simulation J42/G2.

\section{Results}
\label{sec:Results}

\subsection{The flow structure}
\label{subsec:FlowStructure}

\begin{figure*} [htb!]
\centering
\includegraphics[width=0.46\textwidth]{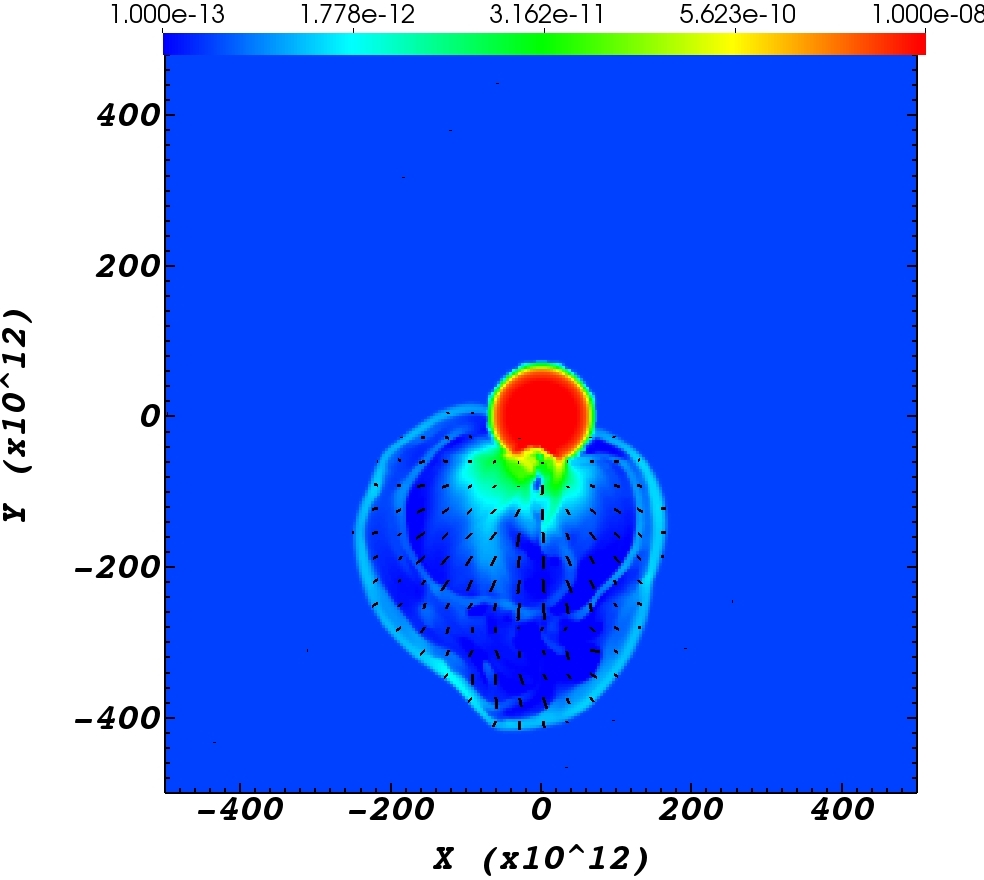}
\includegraphics[width=0.46\textwidth]{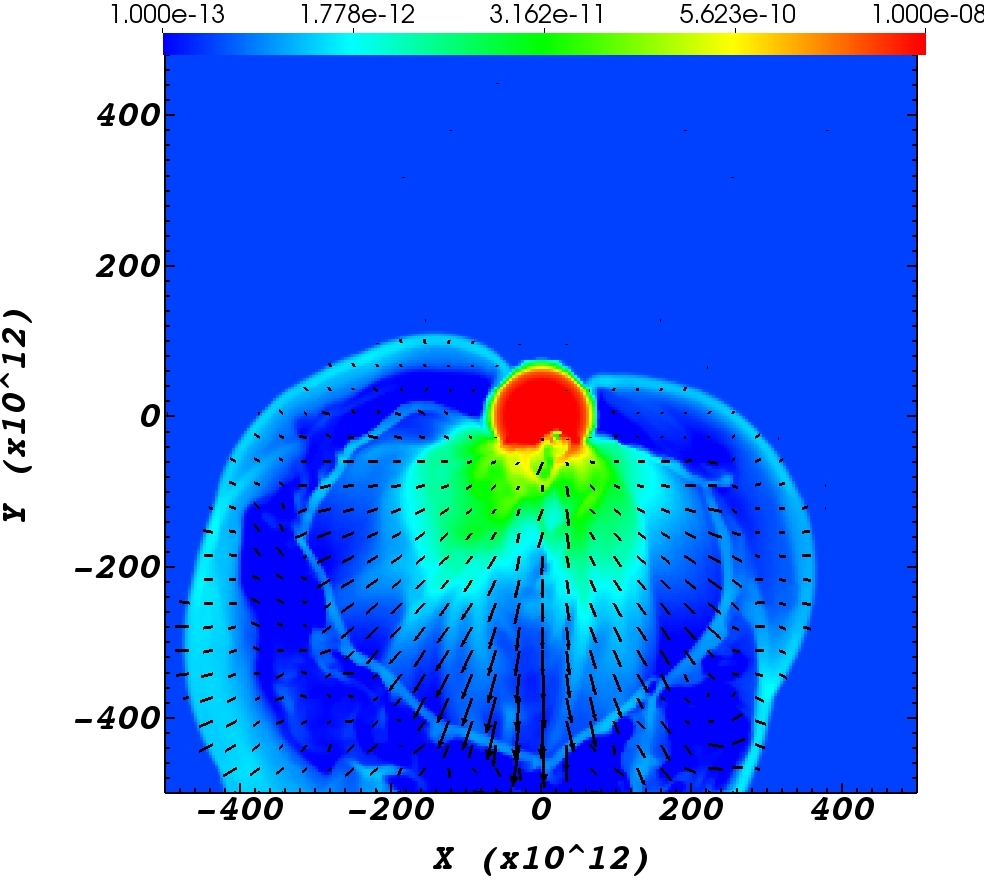}
\includegraphics[width=0.46\textwidth]{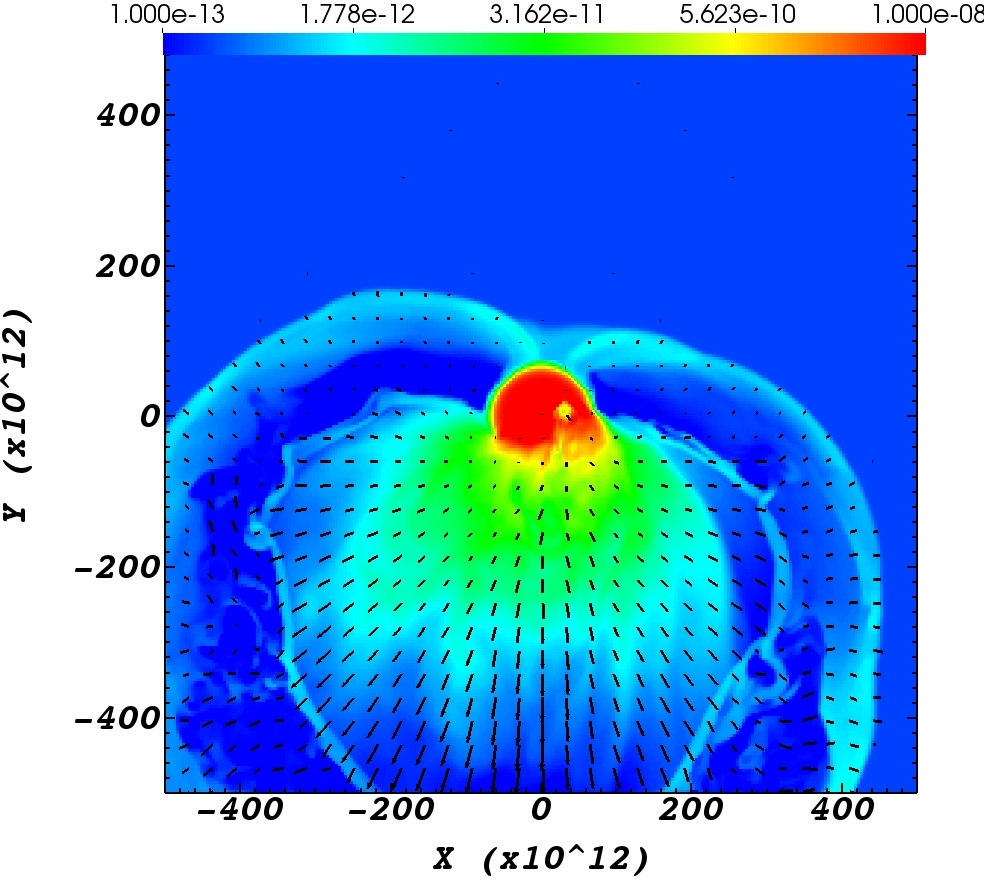}
\includegraphics[width=0.46\textwidth]{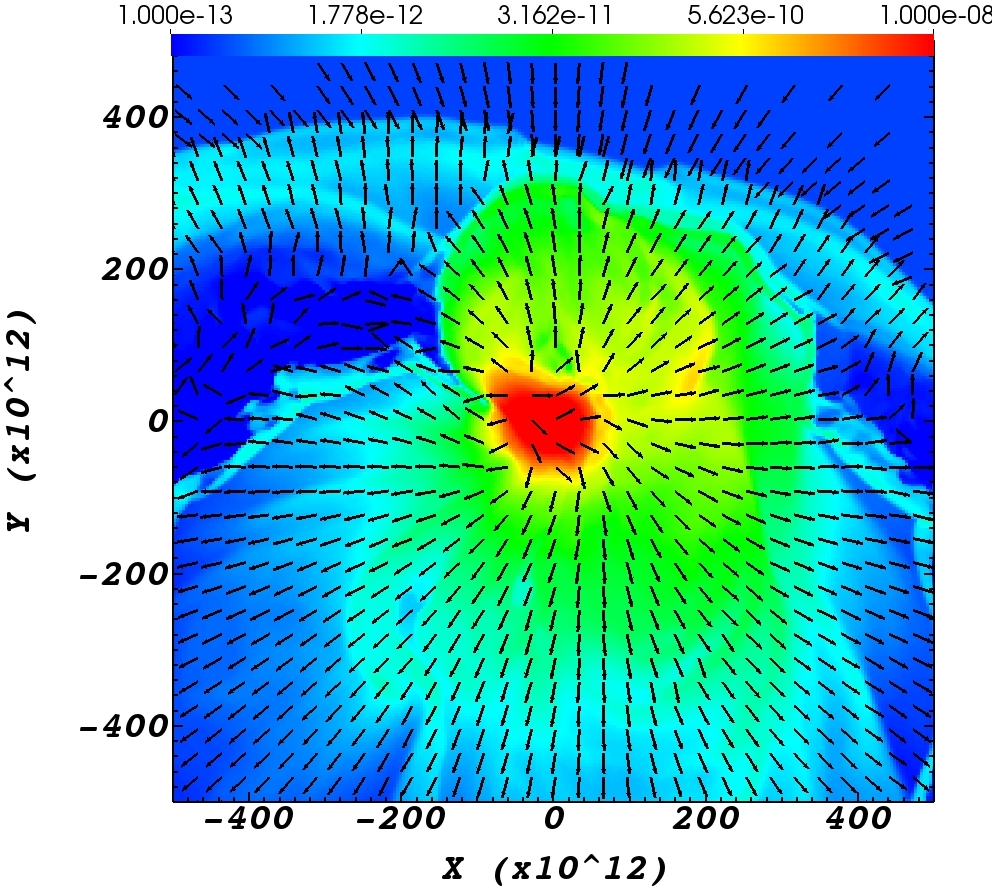}
\includegraphics[width=0.46\textwidth]{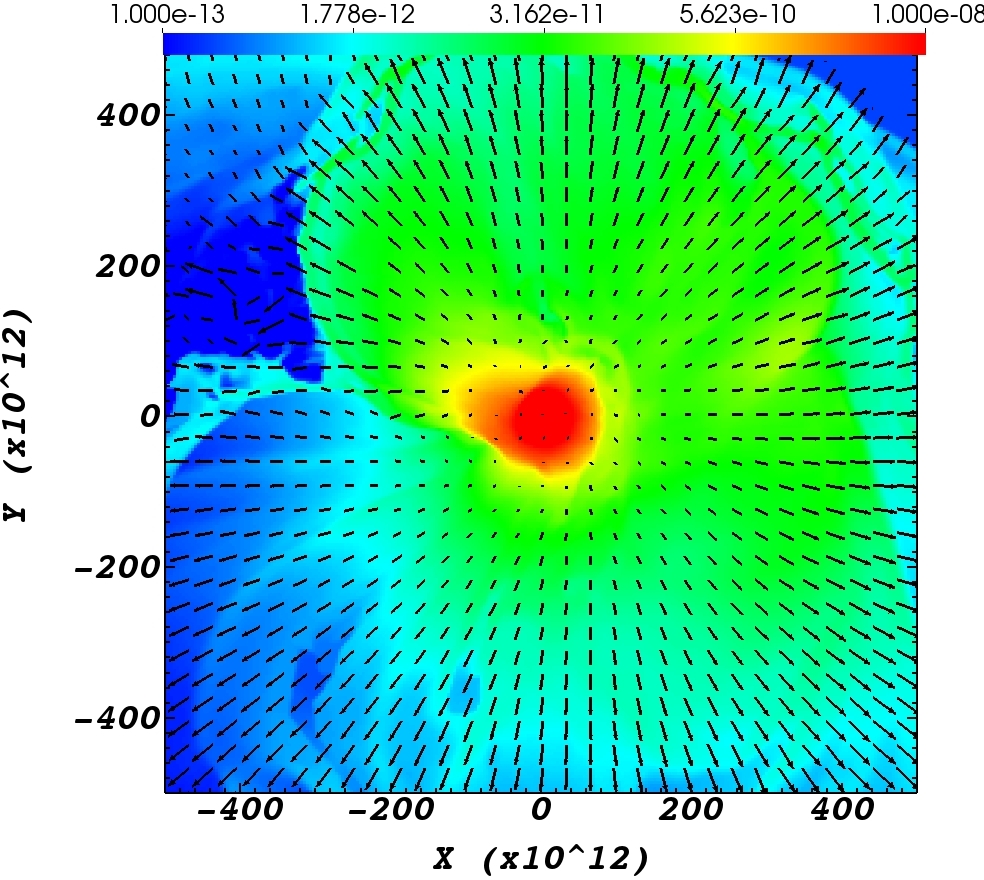}
\includegraphics[width=0.46\textwidth]{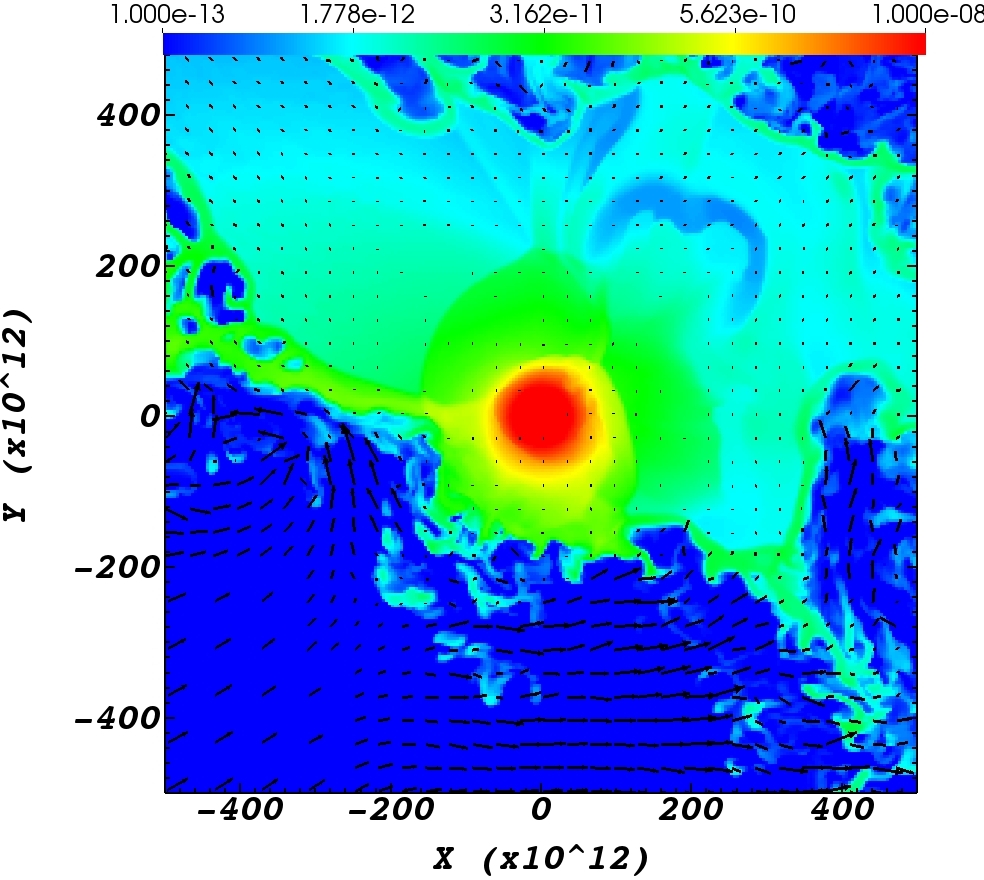}
\caption{Velocity arrows on top of the density maps at $t=2\times 10^6$, $6\times 10^6$, 
$9\times 10^6$, $2.1\times 10^7$, 
$3\times 10^7$ and $1\times 10^8 \sec$, from upper left to lower right, in the equatorial plane of the J42/G2 simulation. 
The lengths of the arrows are proportional to the velocity with a maximum value of $100 \km \s^{-1}$ (all velocities $v>100 \km \s^{-1}$ have the same length as that of $v=100 \km \s^{-1}$). 
The density colour coding is according to the upper colour bar (the same in all panels) and in units of $\g \cm^{-3}$ from $10^{-13} \g \cm^{-3}$ (deep blue) to $10^{-8} \g \cm^{-3}$ (deep red). }
\label{fig:rho_v_XY_fiducial}
\end{figure*}

We here describe the flow structure that results from the jets that the NS launches as it crosses the RSG envelope. We set $t=0$ at the time when the NS enters the RSG envelope. The NS exits the RSG envelope at $t_{\rm jets} = 0.51 \yr$ (out of the $T_{\rm orb}=16.6 \yr$ orbital period), and at that time we stop launching the jets (Fig. \ref{fig:jetspower}). In Fig. \ref{fig:rho_v_XY_fiducial} we present the flow structure of simulation J42/G2 (see Table \ref{Table:cases}). 
In the six panels we present the density maps and velocity directions and magnitudes at six times. In the range of $v=0-100 \km \s^{-1}$ the lengths of the arrows correspond linearly to the velocity, and all velocities of $v>100 \km \s^{-1}$ have the same length as $v=100 \km \s^{-1}$. 

\begin{figure} 
\centering
\includegraphics[width=0.4\textwidth]{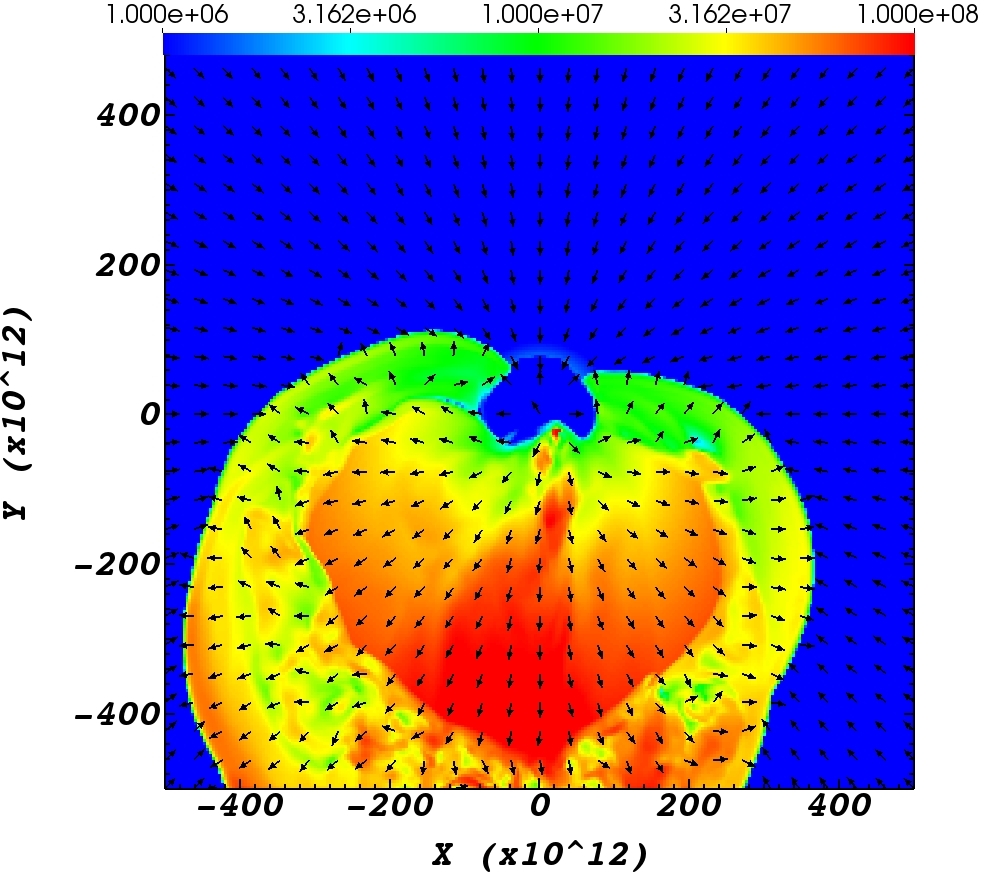} 
\vskip1em
\includegraphics[width=0.4\textwidth]{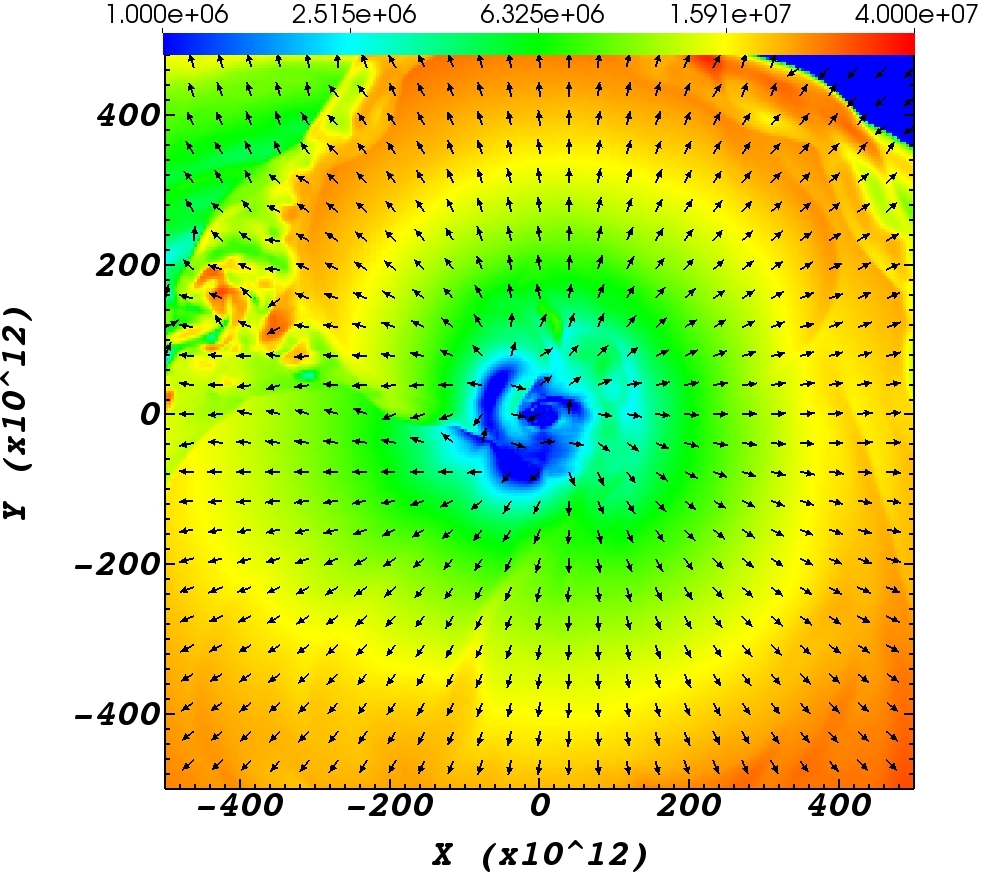}
\caption{Velocity vectors (all arrows of the same length) on top of color-coded velocity maps in the equatorial plane of simulation J42/G2 at $t=0.6\times 10^7 \sec$ (upper panel; deepest red corresponds to velocity of $10^8 \cm \s^{-1}$) and at $t=3\times 10^7 \sec$ (lower panel; deepest red corresponds to velocity of $4 \times 10^7 \cm \s^{-1}$).  
}
\label{fig:Vel_vec_J42}
\end{figure}

\begin{figure*} 
\centering
\includegraphics[width=0.46\textwidth]{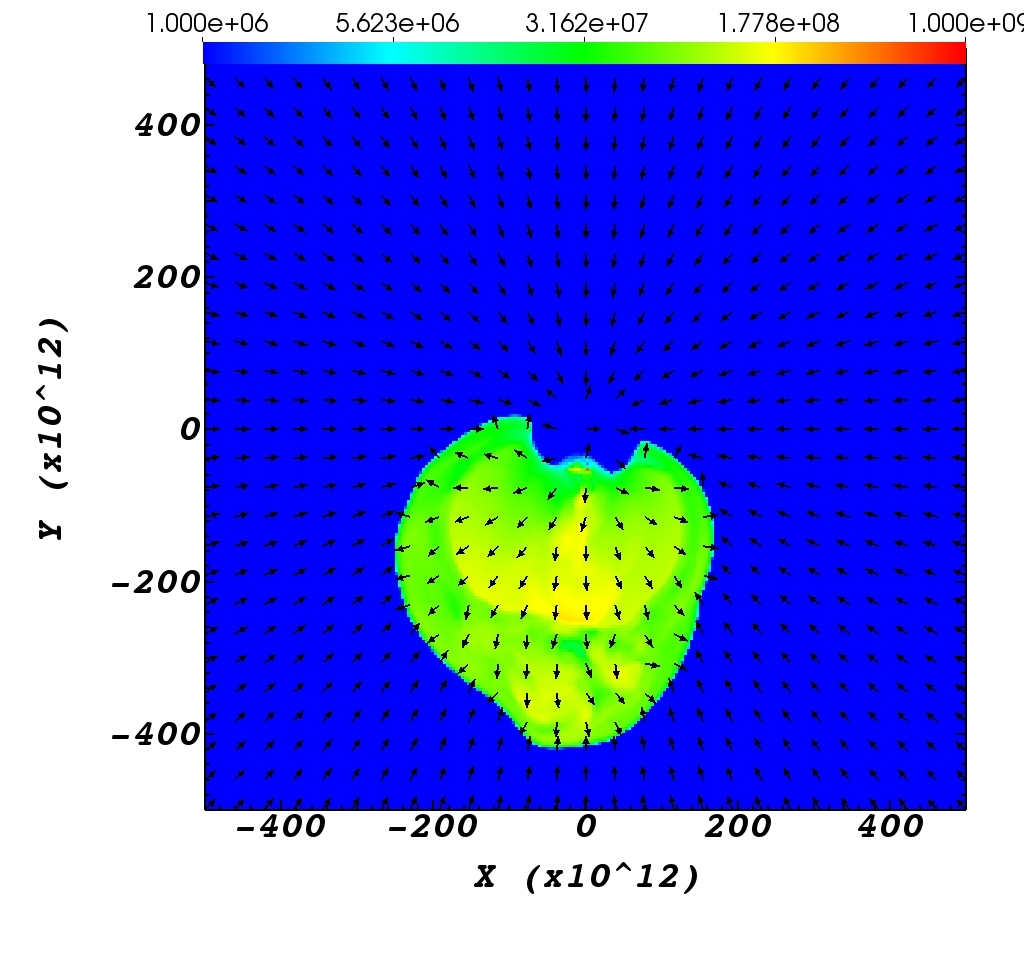}
\includegraphics[width=0.46\textwidth]{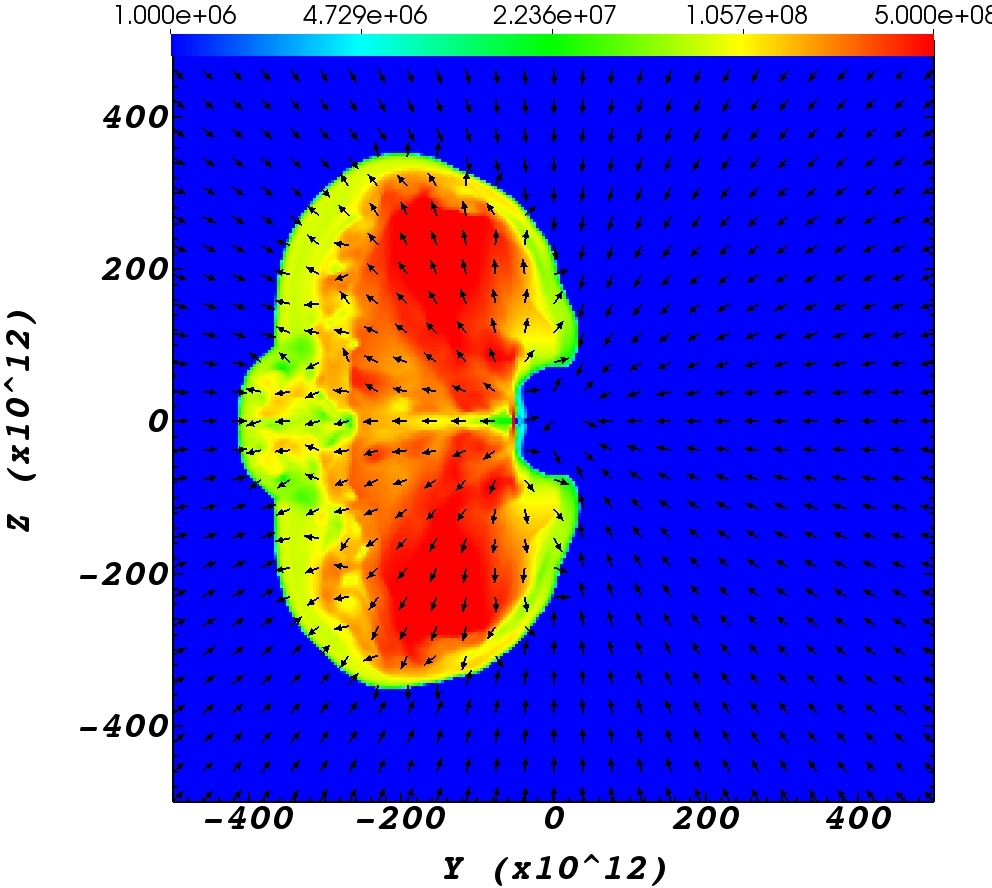}
\includegraphics[width=0.46\textwidth]{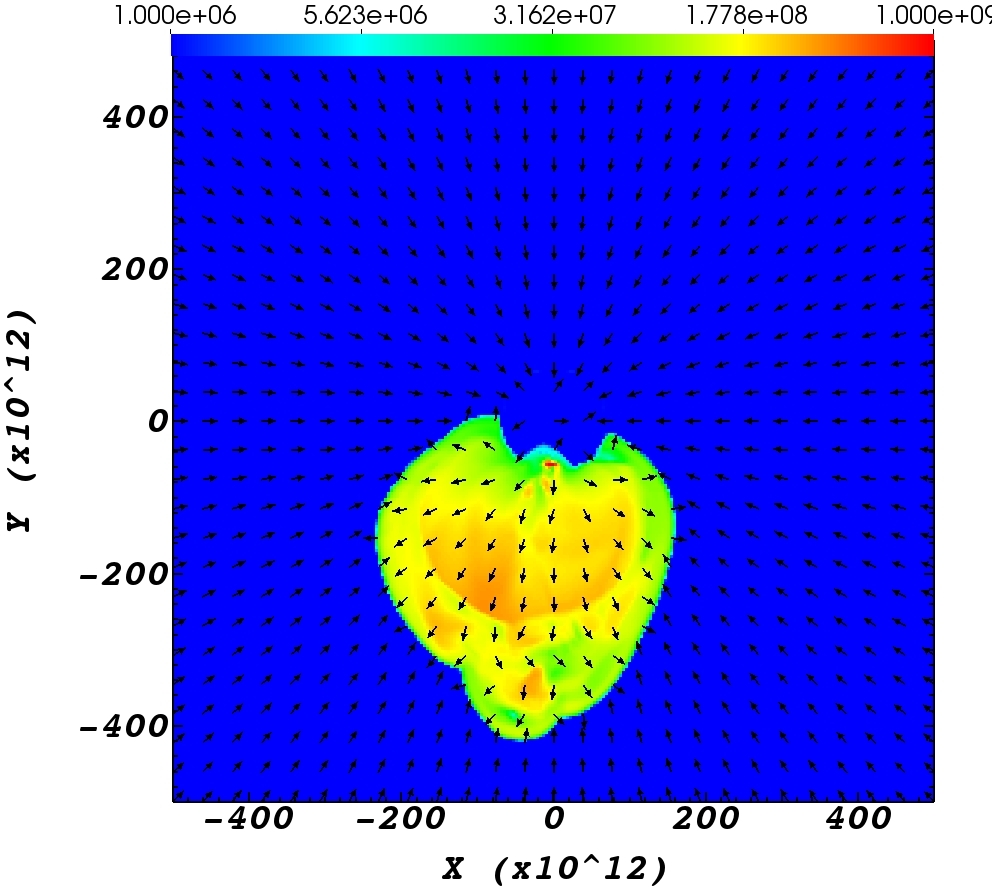}
\includegraphics[width=0.46\textwidth]{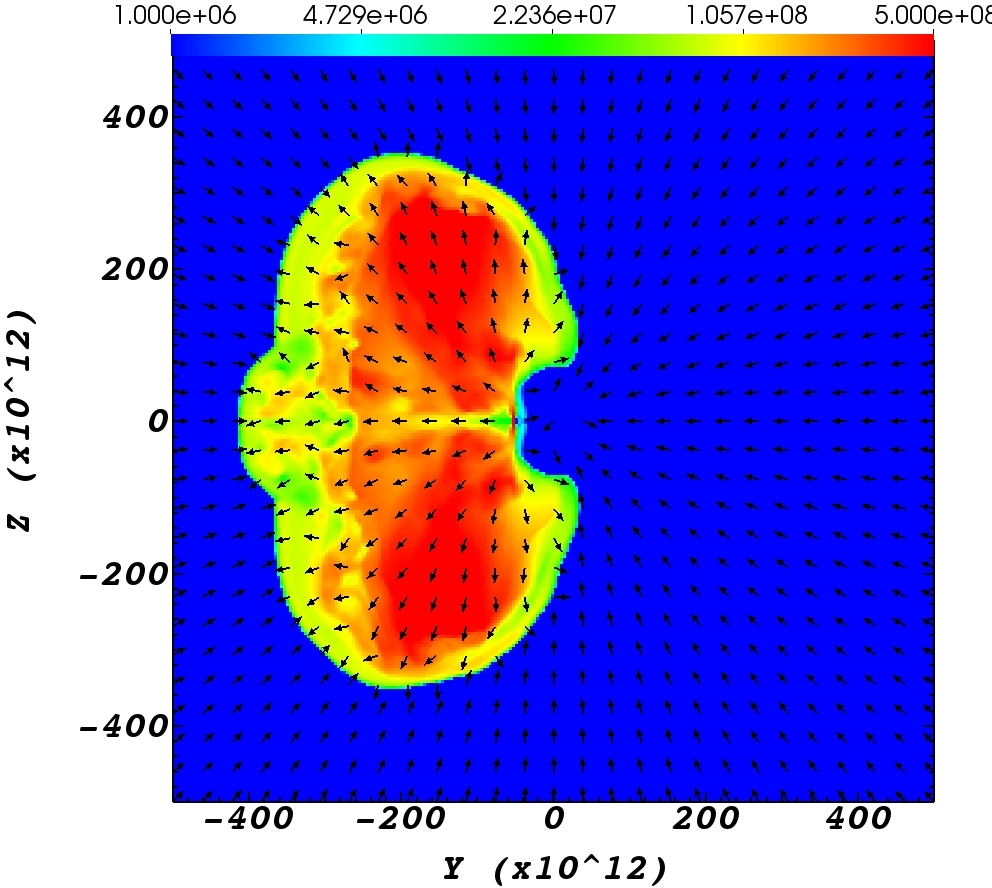}
\caption{Velocity vectors (all arrows of the same length) on top of color-coded velocity maps (in $\cm \s^{-1}$) in the equatorial planes (left panels) and the meridional planes $x=0$ (right panels) of simulation J42/G2 (upper panels) and simulation J43/G2 (lower panels). The upper panels of simulation J42/G2 are at $t=2\times 10^6 \s$  and the lower panels of simulation J43/G2 that has ten times stronger jets are at an earlier time of $1\times 10^6 \s$.  
Deepest red in the left panels corresponds to a velocity of $10^9 \cm \s^{-1}$ and in the right panels to a velocity of $5 \times 10^8 \cm \s^{-1}$} 
\label{fig:J42_vs_J43}
\end{figure*}

To better present the fast outflowing gas we present in Fig. \ref{fig:Vel_vec_J42} the velocity maps in the equatorial plane of simulation J42/G2 at two times. In these maps all arrows are of the same length and the color represent the velocity magnitude with maximum value of $10^4 \km \s^{-1}$ in red. The bipolar jets eject much faster gas in the two polar directions, as we show in the lower panels of Fig. \ref{fig:J42_vs_J43} that present the velocity maps in the meridional plane $x=0$ of two simulations. In Fig. \ref{fig:J42_vs_J43} we also compare simulation J42/G2 (left panels) with simulation J43/G2 (right panels) that has ten times as powerful jets. The left panels of simulation J42/G2 are at $t=2\times 10^6 \s$ and the right panels of simulation J43/G2 are at an earlier time of $10^6 \s$.  As expected, the ejecta reach higher velocities in the more energetic simulation. Some small volumes in the J43/G2 simulation reach velocities that somewhat exceed $10^4 \km \s^{-1}$, but for comparison purposes we limit the red color at $10^4 \km \s^{-1}$.   

From the velocity arrows in Fig. \ref{fig:rho_v_XY_fiducial} and the velocity maps in Figs. \ref{fig:Vel_vec_J42} and \ref{fig:J42_vs_J43} we see that the outflow is not homologous and not radial. The importance of this nonuniform flow is that different parcels of gas collide and convert kinetic energy to thermal energy, some of which ends in radiation (see below).  

In Fig. \ref{fig:P_T_XY_fiducial} we present the pressure (upper row) and temperature (lower row) in the equatorial plane of simulation J42/G2 at two times. We note a shock wave that propagates from the RSG out as a high-pressure high-temperature front (at late times it appears as `arms' inside the grid because the other parts of the shock front are outside the numerical grid). 
We fill the grid with very low density gas at $t=0$ to prevent numerical difficulties. 
The mass that the jets eject out from the envelope collides with this circumstellar matter and drives the shock wave into this gas. 
\begin{figure*} 
\centering
\includegraphics[width=0.46\textwidth]{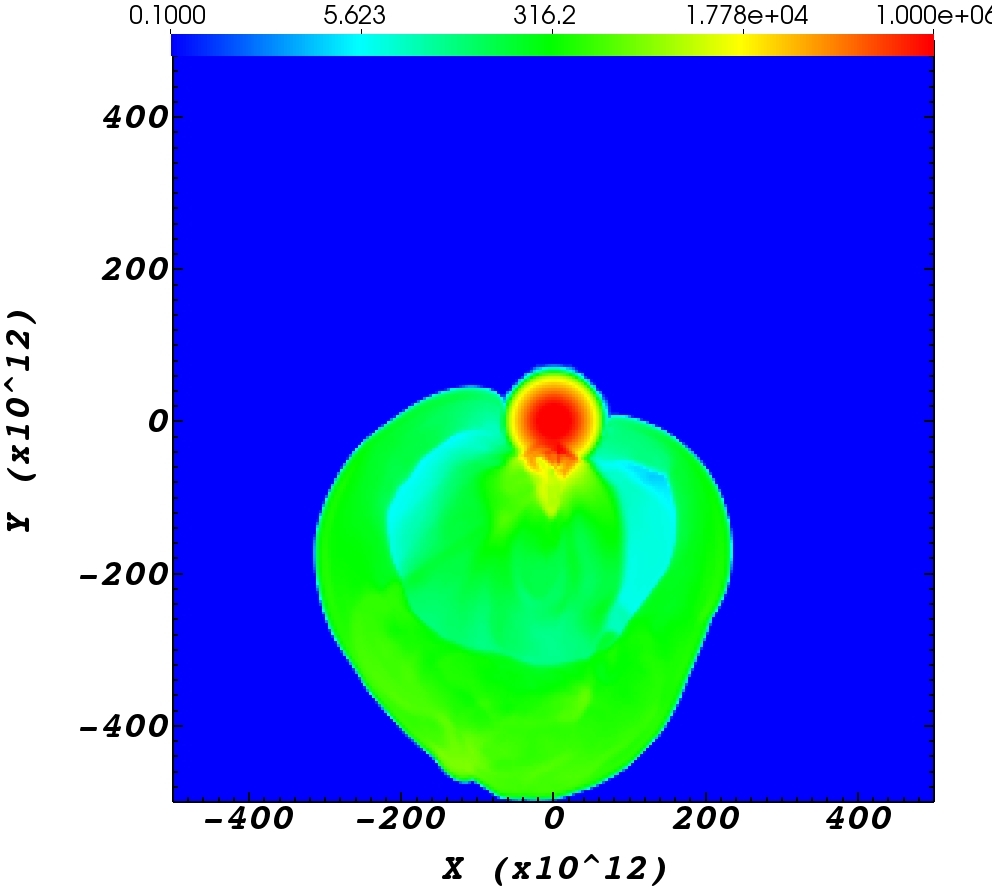}
\includegraphics[width=0.46\textwidth]{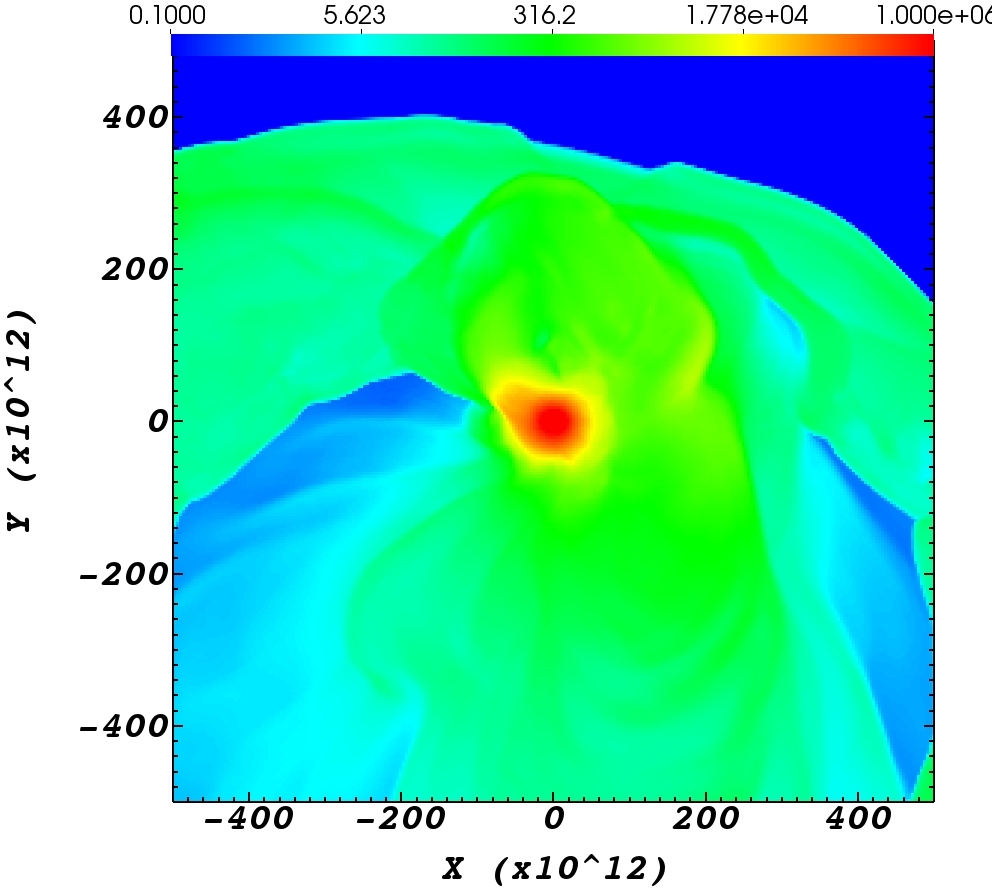}
\includegraphics[width=0.46\textwidth]{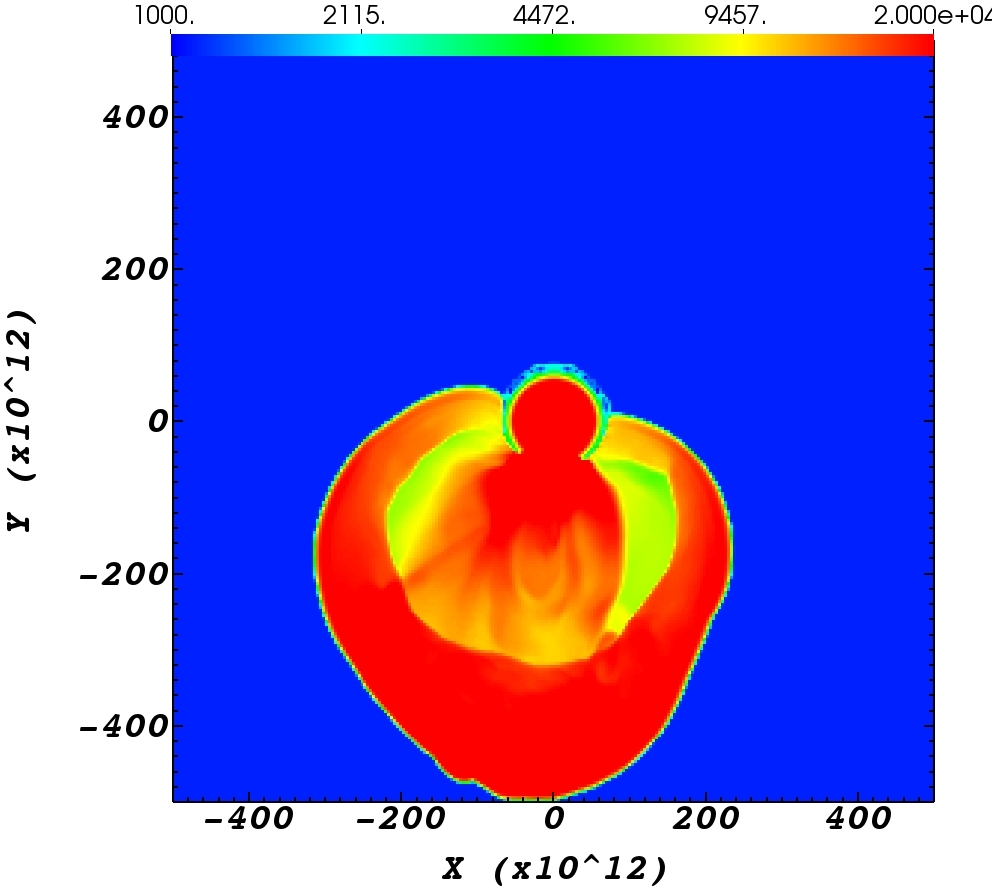}
\includegraphics[width=0.46\textwidth]{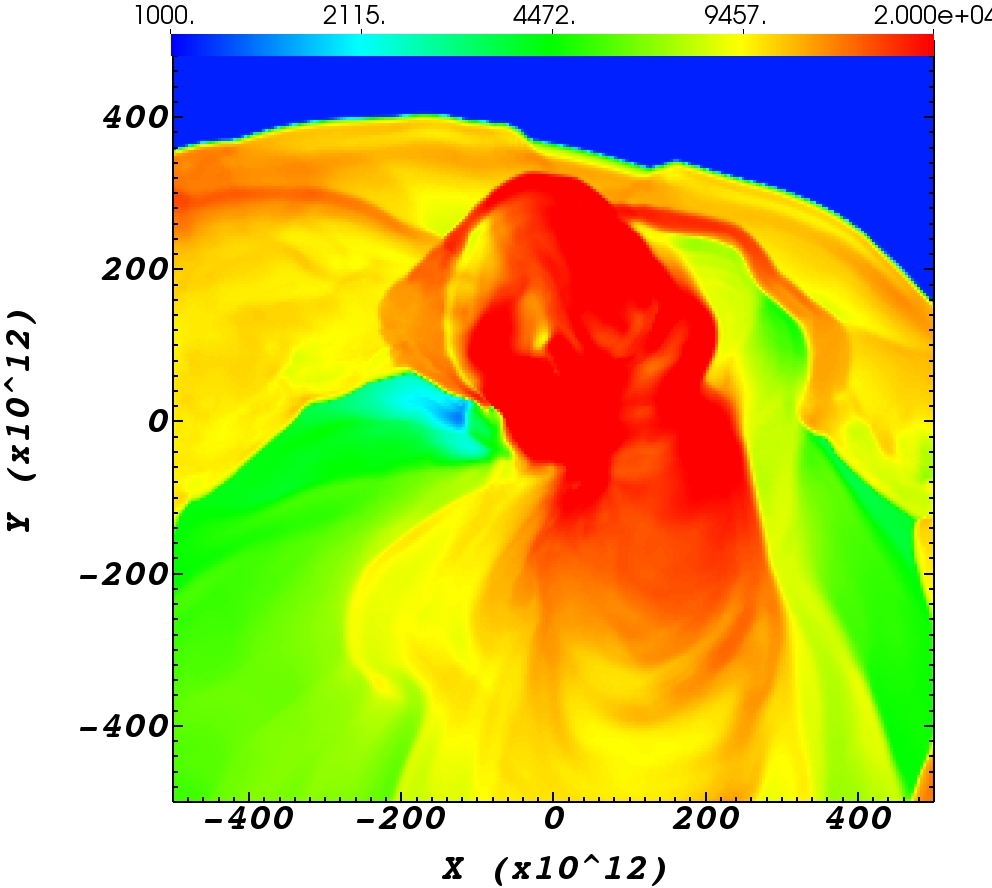}

\caption{Pressure (upper row; deepest red corresponds to $10^6 \erg \cm^{-3}$) and temperature (lower row; deepest red corresponds to $2 \times 10^4 \K$) maps at $t=0.3\times 10^7 \sec$ 
(left column) and at $t=2.1\times 10^7 \sec$ (right column)
in the equatorial plane of simulation J42/G2. 
Pressure and temperature color bars are in $\erg \cm^{-3}$ and $\K$, respectively. 
}
\label{fig:P_T_XY_fiducial}
\end{figure*}
 
As the jets activity continues while the NS gets deeper into the RSG, the jets eject more envelope mass in the general $-y$ direction. New ejecta parts collide with the earlier-ejected mass and the interaction drives shocks into earlier post-shock ejected mass. We see these shocks as high-temperature (red) arcs, one inside the other. This interaction converts kinetic energy to thermal energy. Part of this thermal energy is radiated away (section \ref{subsec:Light Curve}). Later, as the NS is about to exit the RSG envelope the jets eject gas toward the general $+y$ direction (last three panels of 
Fig. \ref{fig:rho_v_XY_fiducial} and right column of 
Fig. \ref{fig:P_T_XY_fiducial}). 

\begin{figure} 
\centering
\includegraphics[width=0.4\textwidth]{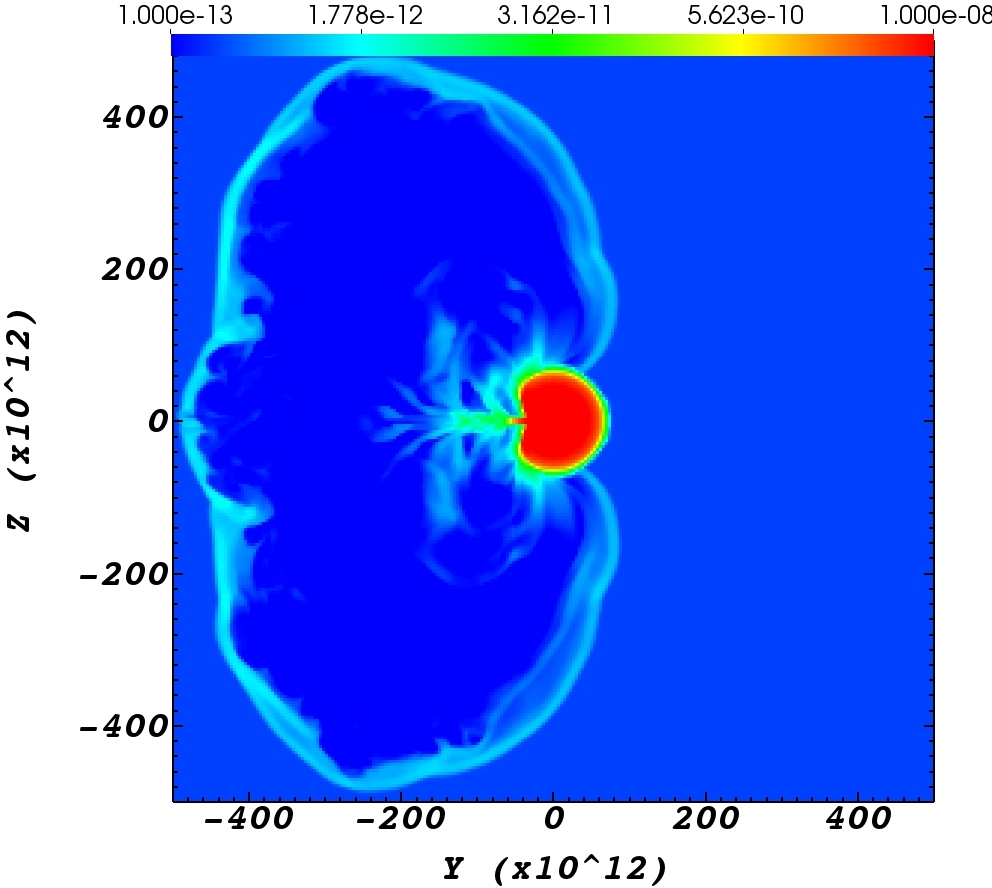}
\includegraphics[width=0.4\textwidth]{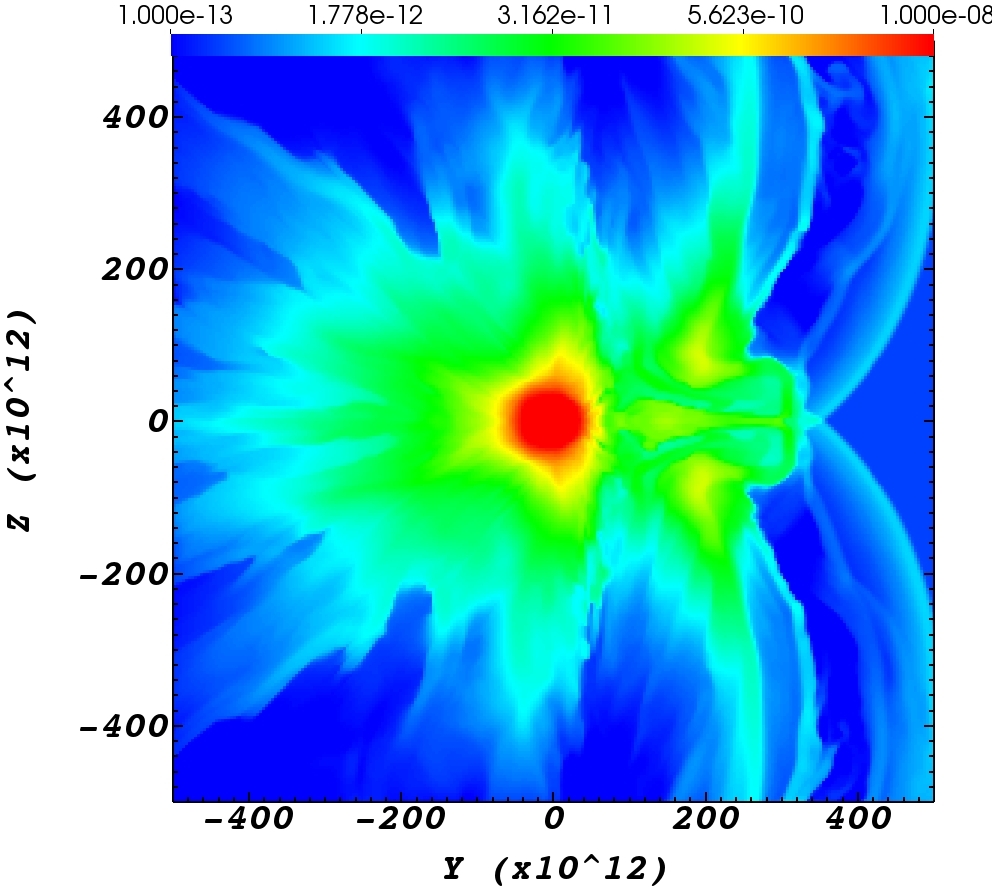}

\caption{Density maps at $t=0.3\times 10^7$ and $2.1\times 10^7 \sec$ 
in the $x=0$ meridional plane of simulation J42/G2. 
The colour coding is the same as in Fig. \ref{fig:rho_v_XY_fiducial} }
\label{fig:rho_v_YZ_fiducial}
\end{figure}

\begin{figure*} 
\centering
\includegraphics[width=0.46\textwidth]{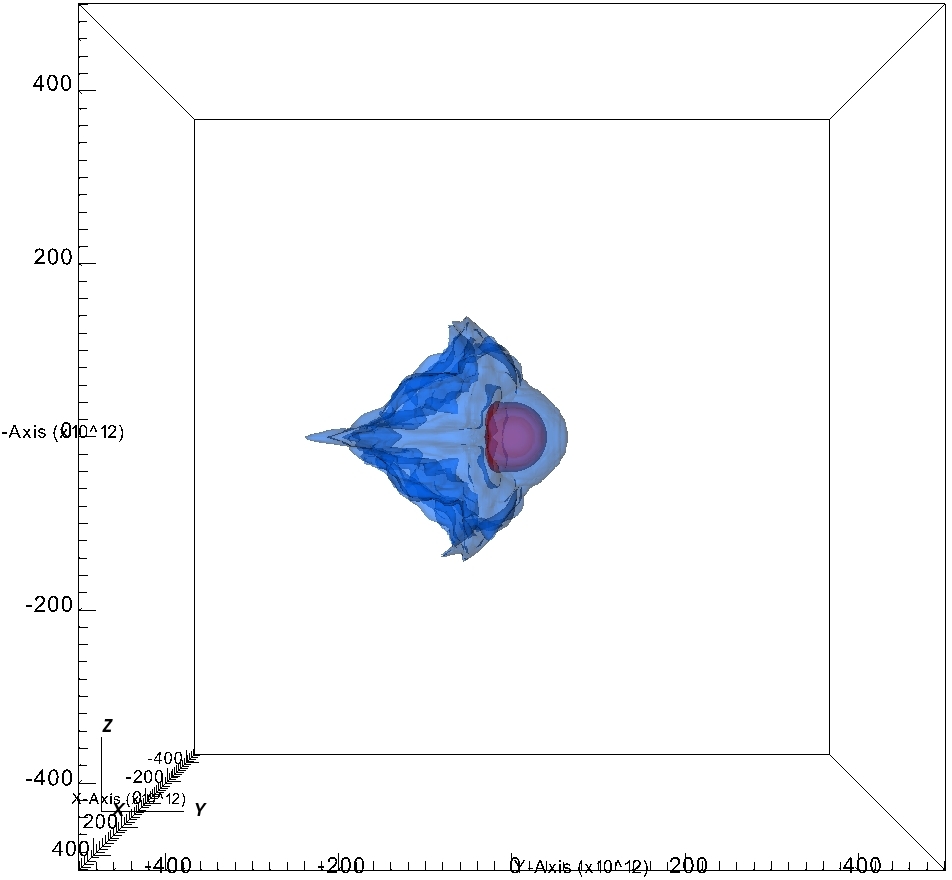}
\includegraphics[width=0.46\textwidth]{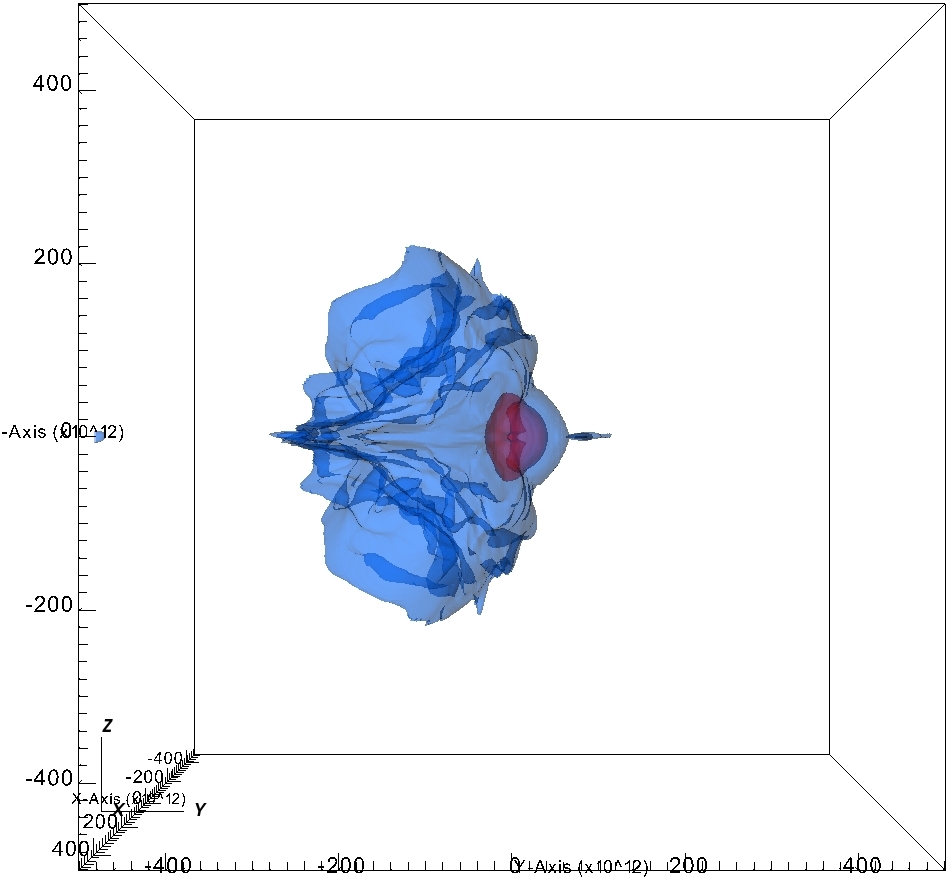}
\includegraphics[width=0.46\textwidth]{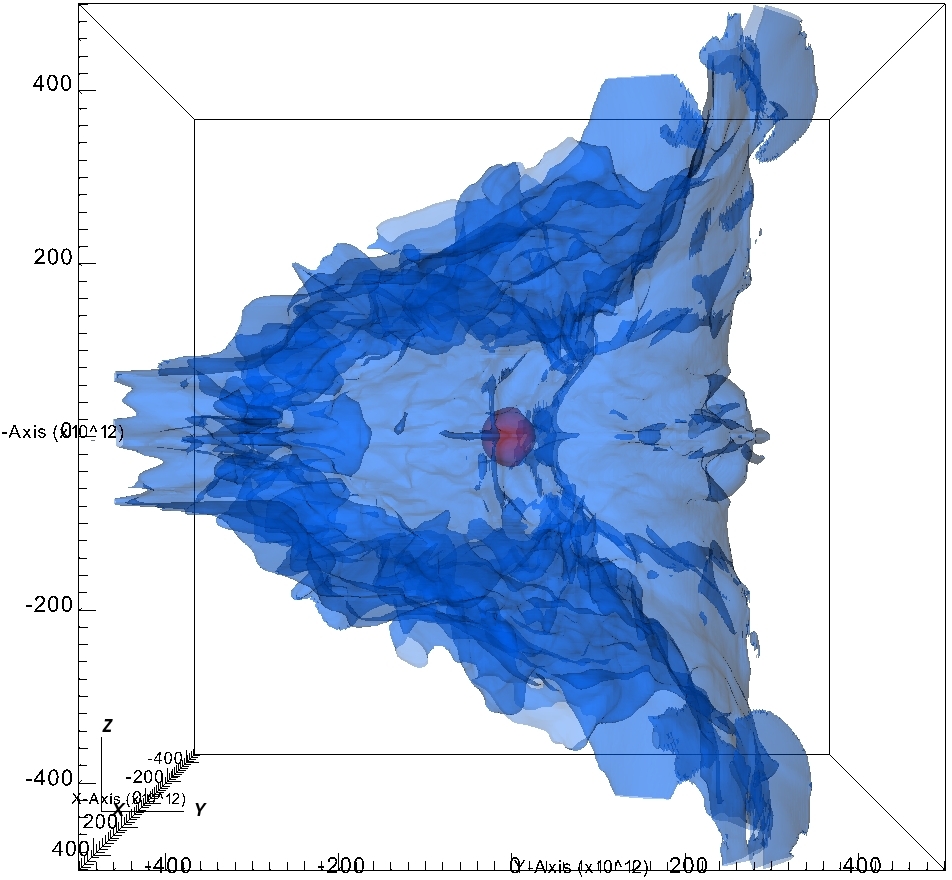}
\includegraphics[width=0.46\textwidth]{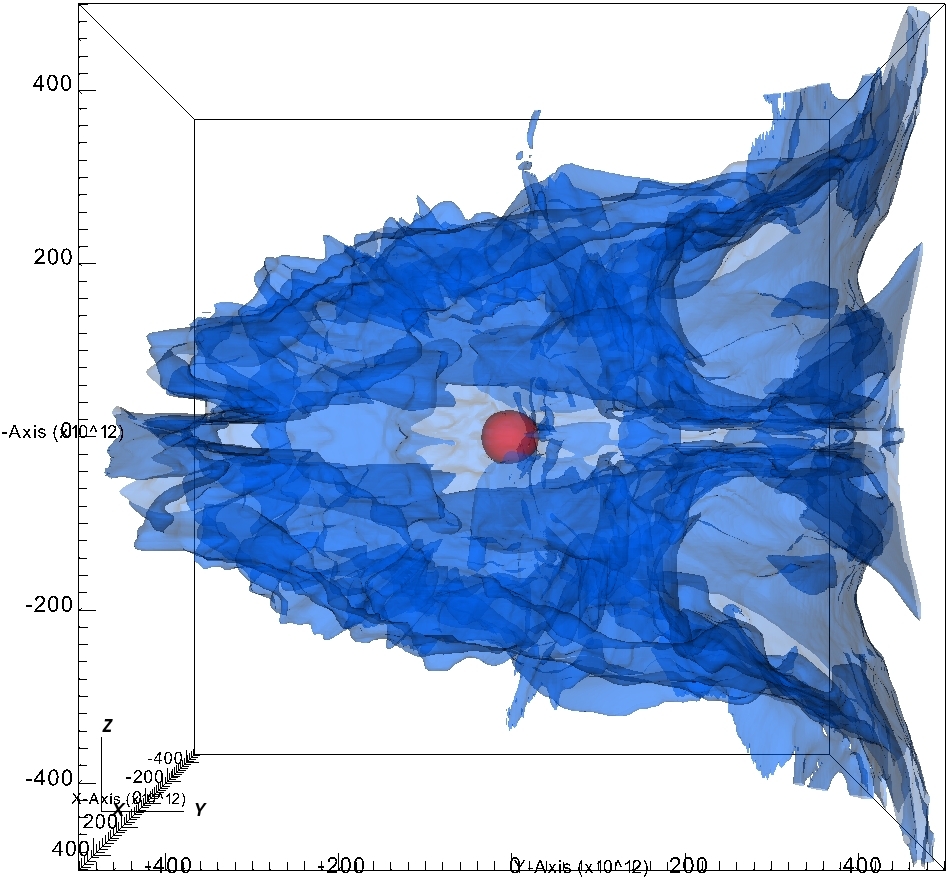}

\caption{Three-dimensional (3D) density maps of simulation J42/G2 at  
$t=0.6\times 10^7 \sec$ (upper left),
$t=0.9\times 10^7 \sec$ (upper right), 
$t=2.1\times 10^7 \sec$ (lower left), 
and  $3.0\times 10^7 \sec$ (lower right). 
The 3D density maps are of two constant-density surfaces, $\rho_{\rm s1} = 10^{-8} \g \cm^{-3}$ (purple) that shows the RSG star vicinity and $\rho_{\rm s2} = 2\times 10^{-12} \g \cm^{-3}$ (blue) that presents the ejecta.
The NS enters the RSG envelope from about the $-y$ direction (left of the figure) and exits in about the $+y$ direction (to the right). 
The jets can be seen in the first two panels as dense (deep red) elongated structures along the $z$ direction (vertical in the figure) inside the RSG envelope.
The RSG star more or less dynamically relaxed to its new structure by the time of the last panel. 
}
\label{fig:3D_fiducial}
\end{figure*}

In Fig. \ref{fig:rho_v_YZ_fiducial} we present the density maps in the meridional plane $x=0$ at two times, of which the early time is somewhat later than the time of the velocity map in the meridional plane that we present in the lower left panel of Fig. \ref{fig:J42_vs_J43}. In Fig. \ref{fig:3D_fiducial} we present three-dimensional (3D) density maps at four different times. The 3D density maps are of two constant-density surfaces, $\rho_{\rm s1} = 10^{-8} \g \cm^{-3}$ that shows the RSG star vicinity and $\rho_{\rm s2} = 2 \times 10^{-12} \g \cm^{-3}$ that depicts the ejecta.  Figs. \ref{fig:rho_v_YZ_fiducial} and \ref{fig:3D_fiducial} further emphasise the complicated large-scale outflow morphology and its clumpy structure. 

Not unexpectedly, we learn from the results we present in this section that the jets inflate the envelope mainly in the general directions where the NS enters and exits the envelope, including mass flowing at large angles to the equatorial plane when the jets are powerful (more in the next section). 
This morphology results from that when the jets are deep inside the envelope they inflate the inner envelope. The pronounced inflation of the outer envelope occurs when the NS launches the jets in the envelope outskirts, namely, where it enters and where it exits the envelope. 
Another important result is that the ejecta is clumpy and within the boundary of our grid does not reach yet homologous expansion.

\subsection{Outflow geometry}
\label{subsec:OutfloeGeometry}
 
\begin{figure*} 
\centering
\includegraphics[width=0.46\textwidth]{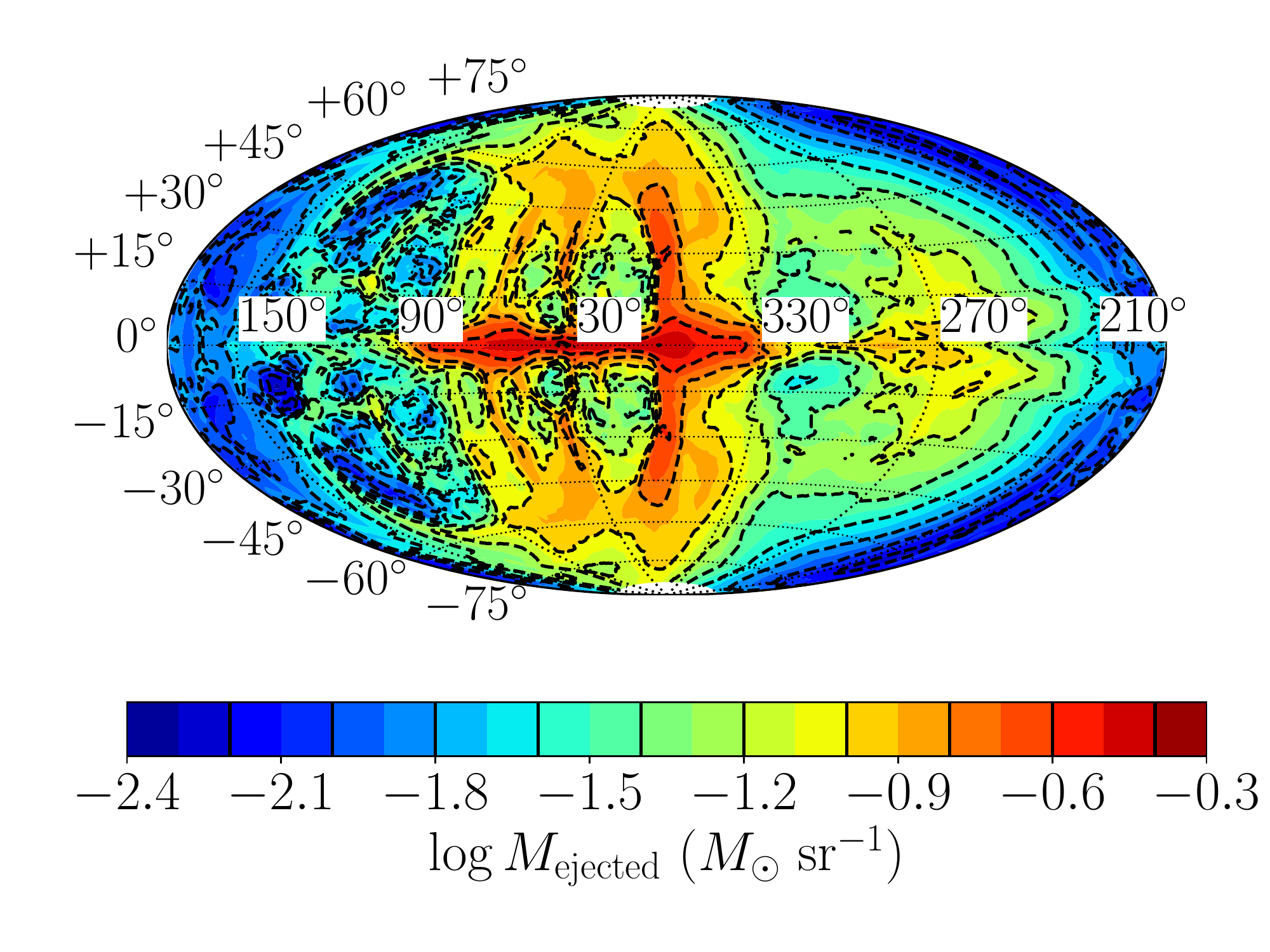}
\includegraphics[width=0.46\textwidth]{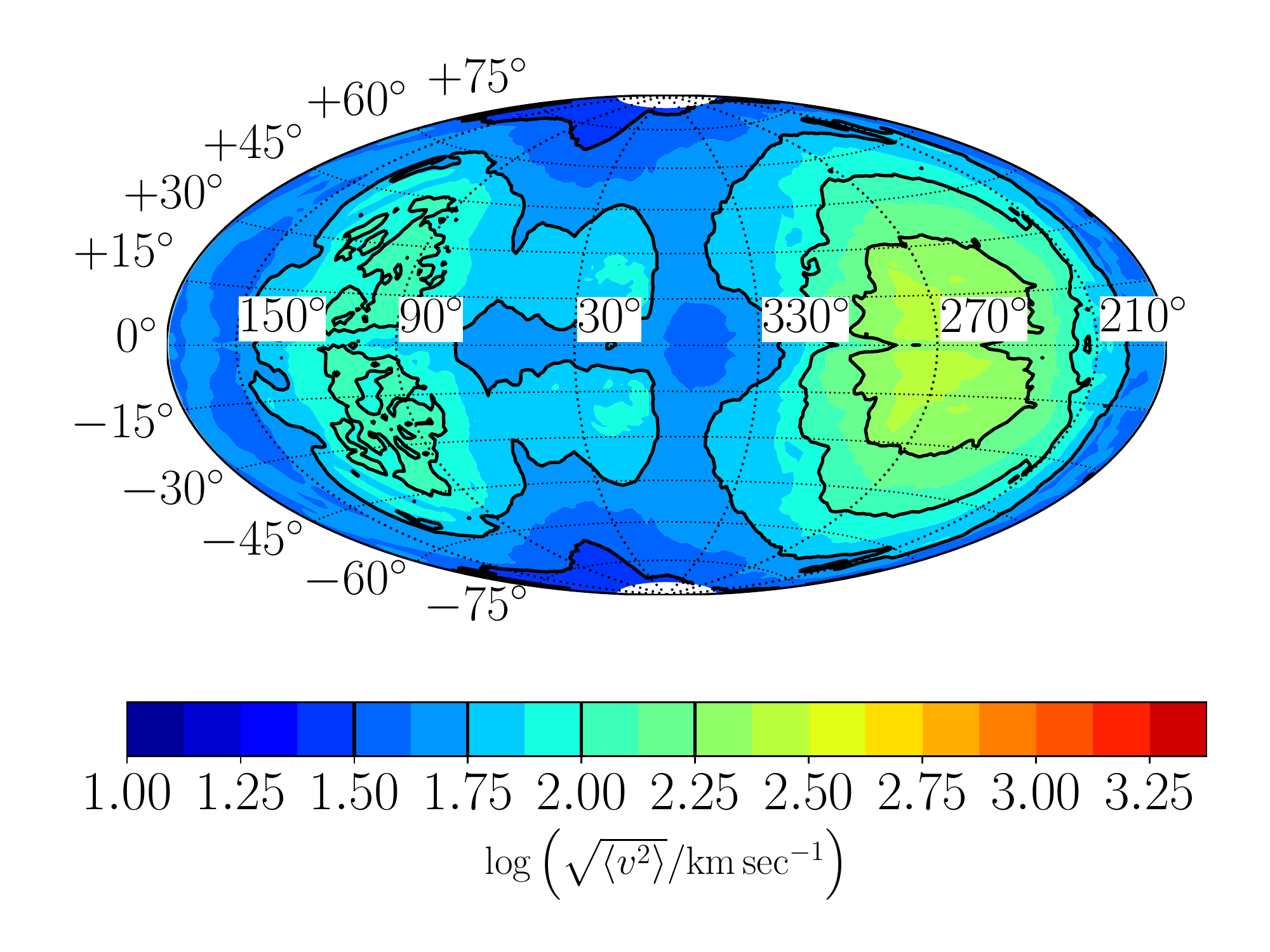}
\includegraphics[width=0.46\textwidth]{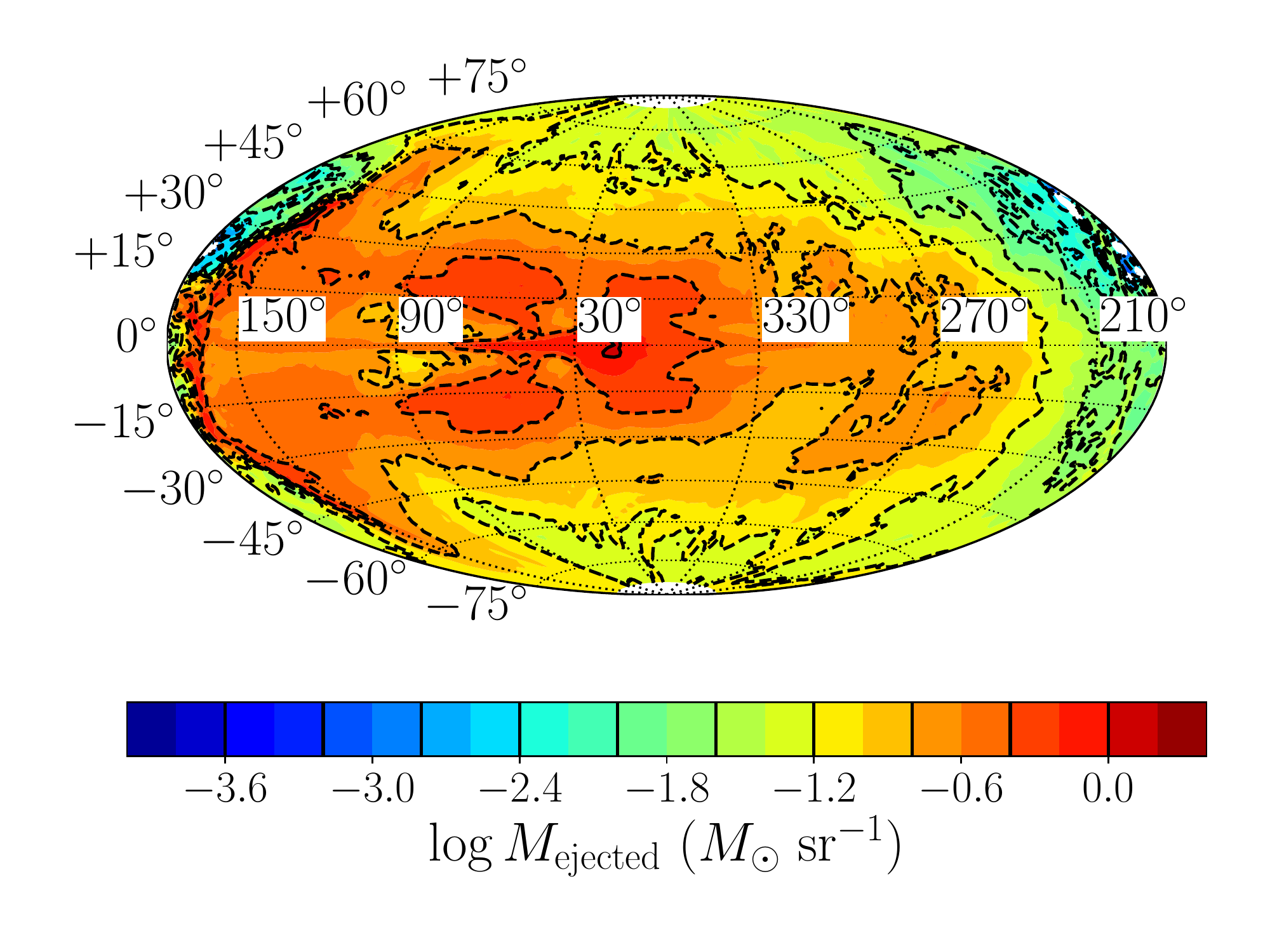}
\includegraphics[width=0.46\textwidth]{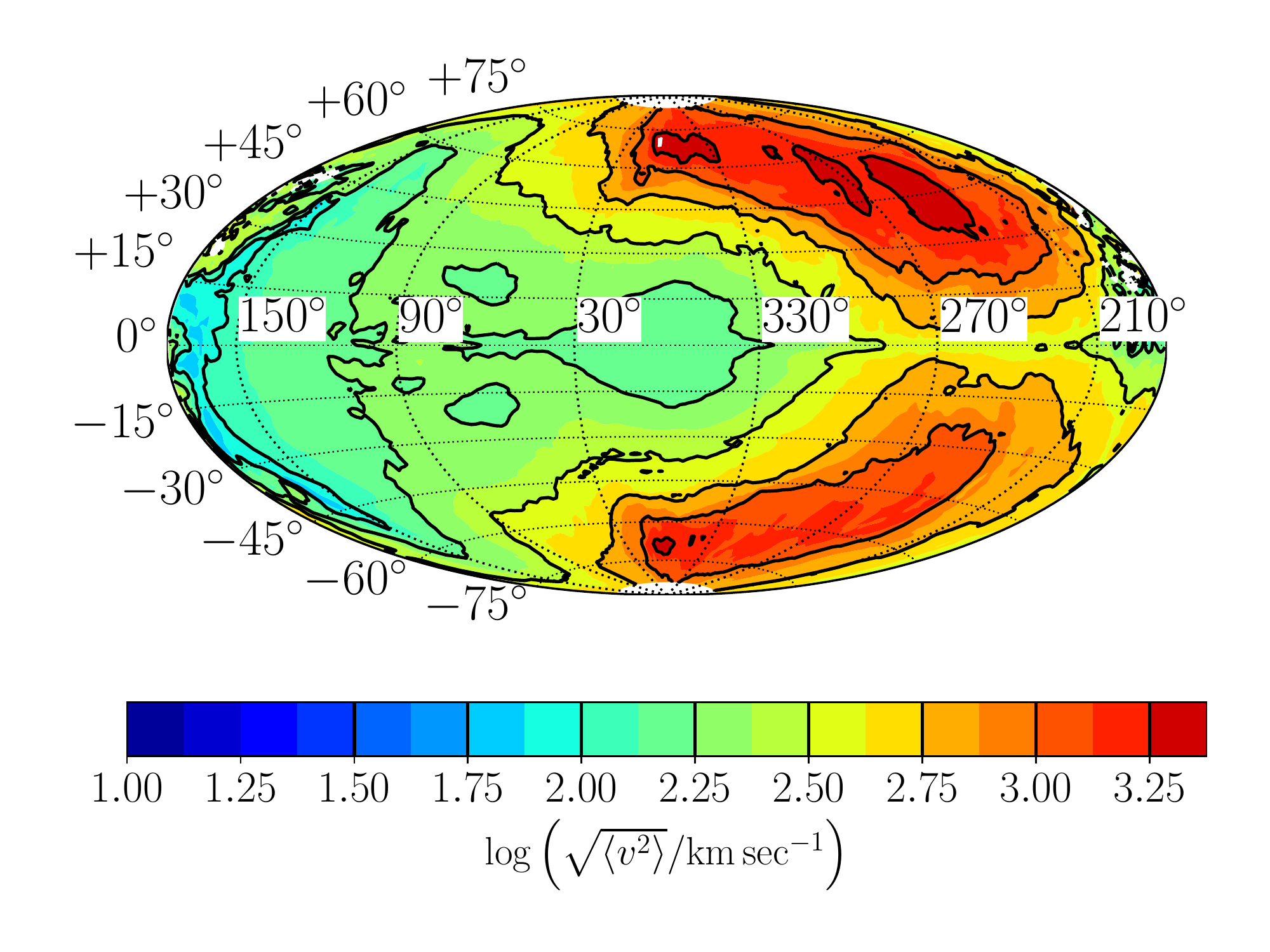}
\caption{Upper panels: The total outflow mass per units solid angle (left) and $v_{\rm rms} (\theta,\phi) \equiv \sqrt{<v^2>}$ (right) for simulation J41/G1. Lower panels: Similarly but for simulation J42/G2. The latitude $\theta=0$ coincides with the equatorial plane and the NS enters the RSG envelope from the longitude $\phi \simeq 270^\circ$ direction and exits from the longitude $\phi \simeq 90^\circ$ direction. 
}
\label{fig:outflow_maps}
\end{figure*}

In Fig. \ref{fig:outflow_maps} we present the total (integrated over the time of our simulation of $10^8 \s$) outflow mass per unit solid angle as function of the direction $M_{\rm ejecta} (\theta,\phi)$ (left panels), and the average velocity calculated from the kinetic energy outflow $v_{\rm rms} (\theta,\phi) \equiv \sqrt{<v^2>} = \sqrt{2 E_{\rm k} (\theta,\phi)/M_{\rm ejecta} (\theta,\phi)}$, where $E_{\rm k} (\theta,\phi)$ is the total kinetic energy outflow per units solid angle (right panels). In the upper row we present the results for simulation J41/G1 and in the lower panel for simulation J42/G2. The latitude $\theta=0$ is in the equatorial plane and the longitude $\phi=0$ corresponds to the $+x$ direction. The NS enters the RSG envelope from the $\phi \simeq270^\circ$ direction and exits from the $\phi \simeq 90^\circ$ direction.
By `outflow' we refer here to all mass, bound and unbound, that crosses a sphere of radius $R_{\rm out}=2.5 \times 10^{15} \cm$ in simulation J41/G1 and of radius $R_{\rm out}=5 \times 10^{15} \cm$ in simulation J42/G2. The majority of this gas is unbound as we can infer from the velocity maps (section \ref{subsec:FlowStructure}) as the velocities are much larger than the escape velocity of $26 \km \s^{-1}$ at $r=5 \times 10^{14} \cm$. As we mentioned in section \ref{subsubsec:spherical}, for a test simulation without the jets there is a weak inflow of mass into the numerical grid. For that, our calculated outflow mass is not influenced by more than several per cents because of the departure from exact hydrostatic equilibrium of the initial stellar model. In any case, we do not overestimate the outflowing mass. 

\begin{figure} 
\centering
\includegraphics[width=0.45\textwidth]{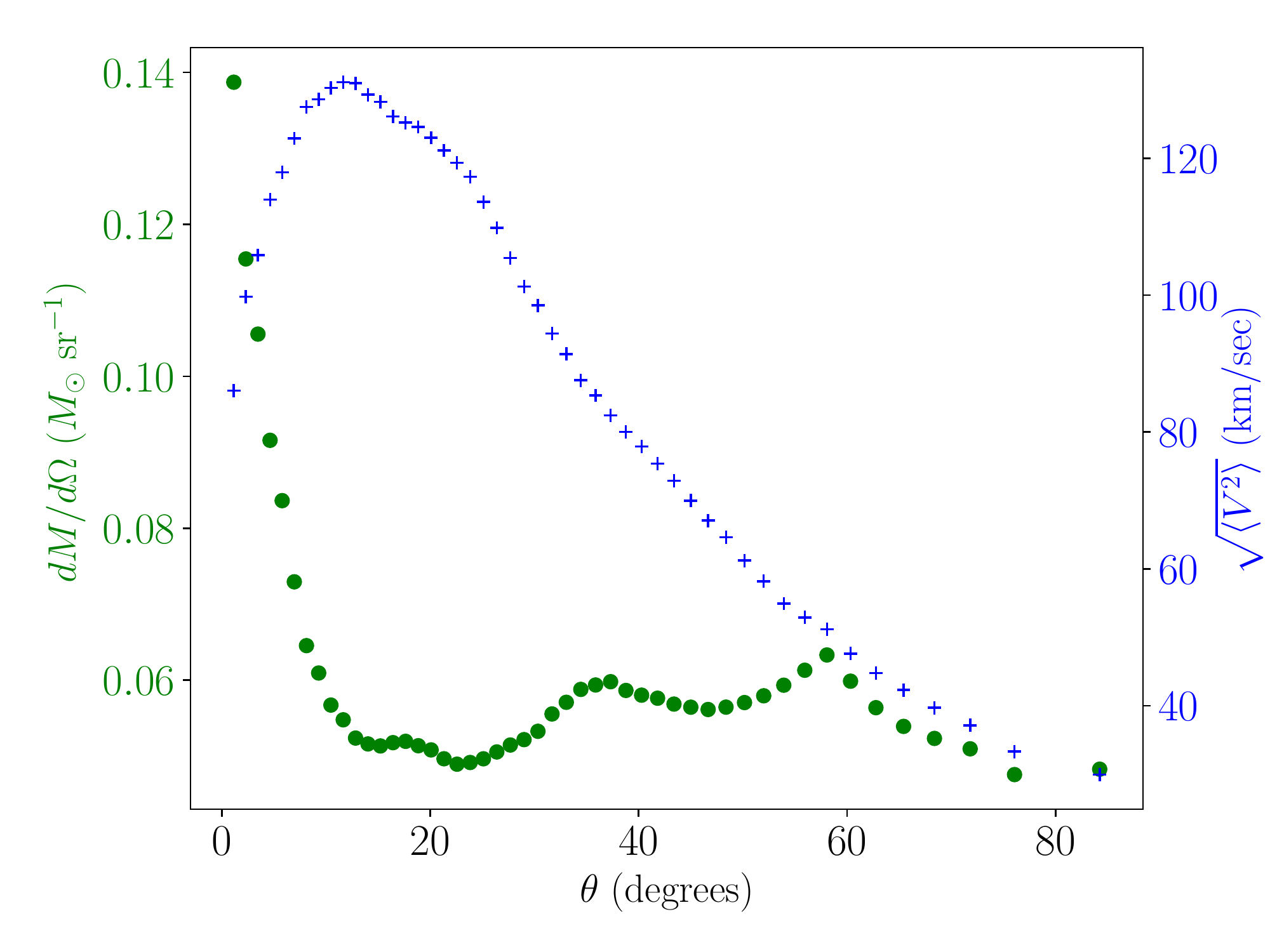} \\ 
\includegraphics[width=0.45\textwidth]{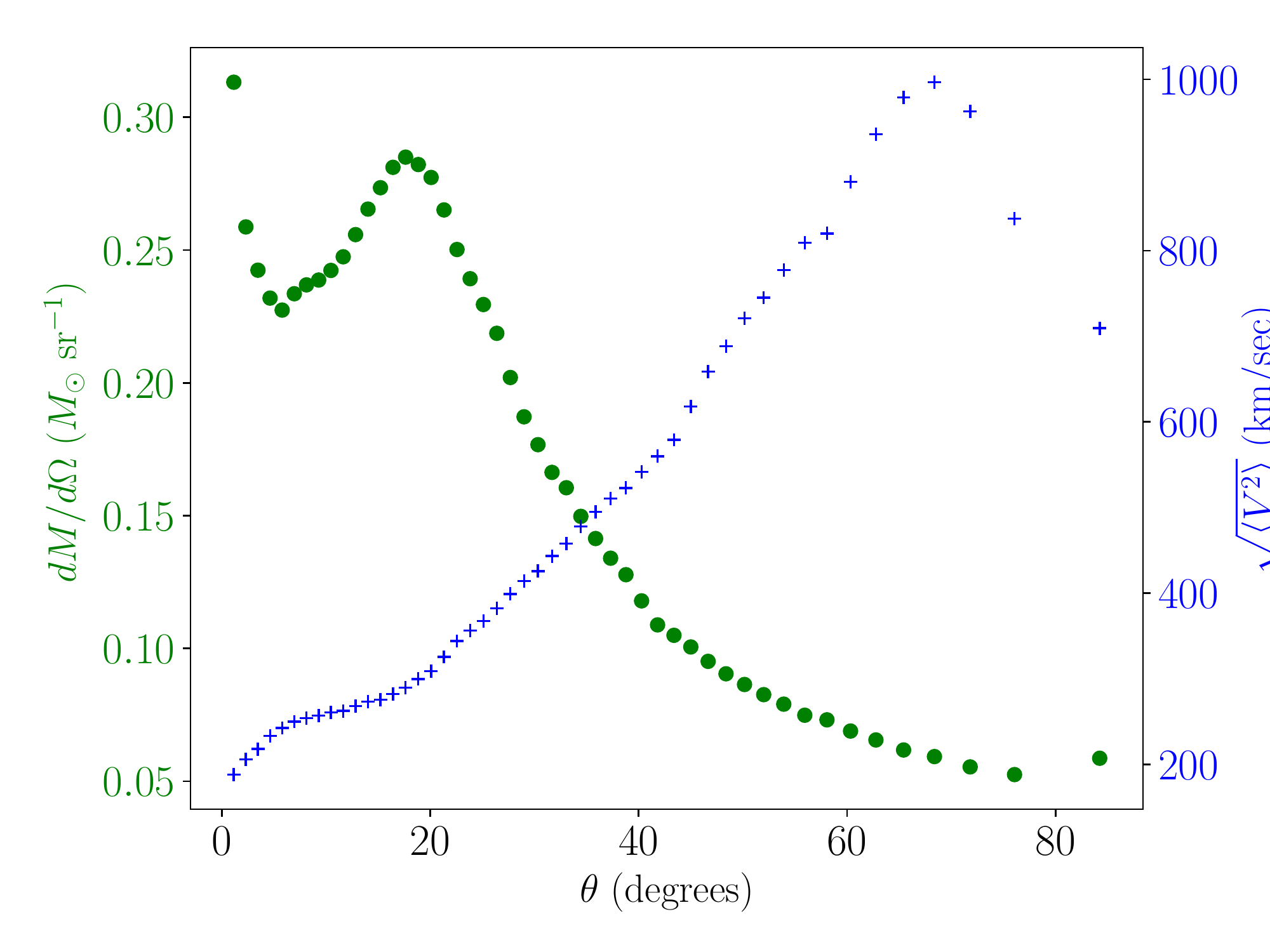}

\caption{Quantities averaged over all longitudes and over the two sides of the equatorial plane. 
The green-dots lines present $M_{\rm ejecta} (\theta)$ which is the average of $M_{\rm ejecta} (\theta, \phi)$ and the blue-pluses lines present the average of $v_{\rm rms} (\theta,\phi)$. 
Upper panel is for simulation J41/G1 and the lower panel is for simulation J42/G2. 
}
\label{fig:outflow_theta}
\end{figure}

In Fig. \ref{fig:outflow_theta} we present the average over all longitudes (from $\phi=0$ to $\phi=360^\circ$) of the same quantities that we present in Fig. \ref{fig:outflow_maps}. We also average over the two sides of the equatorial plane, i.e., we group the $-\theta$ direction with the $+\theta$ direction.  

Figs. \ref{fig:outflow_maps} and \ref{fig:outflow_theta} show that there are large qualitative differences between the outflow morphology of simulations J41/G1 and J42/G2. In the J41/G1 case the largest mass concentration is near the equatorial plane at longitudes from $\phi \simeq 300^\circ (= - 30 ^\circ)$ to $\phi \simeq 90^\circ$, with high mass concentration to large latitudes at around $\phi \simeq 15^\circ$. 
In the ten times as powerful jets case J42/G2, on the other hand, the jets accelerate the high outflow mass rate to an extended solid angle in the forward direction (the direction to which the NS exits the envelope), and around $(\theta, \phi) \simeq (0,90^\circ)$. The green-dots lines in Fig. 
\ref{fig:outflow_theta} also demonstrate the differences in the outflowing mass morphologies between the two simulations. 

The morphology of the fastest outflowing gas is also different. In simulation J41/G1 there are two general directions, $\phi \simeq 120^\circ$ and $\phi \simeq 270^\circ$, of fast outflowing gas with average velocities of $\simeq 200 \km \s^{-1}$ (green zones). Both regions show the bipolar structure that the jets induce. The bipolarity appears as fast outflowing directions above and below the equatorial plane.  
The case with simulation J42/G2 is different. Here the fastest average outflowing gas reach velocities of $\simeq 2000 \km \s^{-1}$ and much closer to the poles ($\theta = \pm 90^\circ$), and only in the $\phi \simeq 270^\circ$ direction, which is about the longitude from which the NS enters the RSG envelope, but not from the $\phi \simeq 90^\circ$ direction. 
The peak at $60^\circ \la \vert \theta \vert \la 80^\circ$ in the blue-pluses line in the right panel of Fig. \ref{fig:outflow_theta} also demonstrates the bipolar structure of the fast outflow, to both sides of the equatorial plane, 

The main points to take from the discussion of the outflow morphologies are the following. 
(1) The outflow morphology, i.e., the mass outflow rate and velocity as functions of direction, depends on the jets' power and can vary a lot between different cases. 
(Future studies should examine the dependence of the outflow morphologies on the manner by which one injects the jets into the numerical grid.) 
(2) The outflow morphology has only a mirror symmetry about the equatorial plane. There are no other symmetries, i.e., nor axi-symmetry, nor simple dipole, and nor simple bipolar structures. (3) As we discussed in section \ref{subsec:FlowStructure}, the outflowing gas is clumpy with relatively large variations in the outflow velocities in different directions.  

\subsection{Light Curve}
\label{subsec:Light Curve}

Because of the very complicated structure of the outflow (ejecta) and the density fluctuations of the ejecta, 
we can only very crudely estimate the light curve. 
For example, because of the highly non-spherical outflow and density distribution the light curve will depend also on the relative direction of the observer.  
Another difficulty is that in some directions ($-y$ and $+y$) during part of the time the photosphere is outside the numerical grid. As a result of these limitations we can estimate only the general properties of the expected light curve.

Along the axes the large grid (G2) extends to $5\times 10^{14} \cm$. Ejecta densities at this distance from the centre along and near the $x=0$ axis, where most ejecta leaves the grid, are (Fig. \ref{fig:rho_v_XY_fiducial}) $\approx 10^{-12} \g \cm^{-3}$. For an opacity of $\simeq 0.1$ the optical depth there is $\tau \approx 50$. In other directions the optical depth is lower even at that distance. 
The smaller grid extends to half this distance, but the G1 simulations have lower ejecta densities and so optical depth is lower. 
Overall, the location of the highly-non-spherical photosphere of the hot ($\approx 10^4 \K$) ejecta is crudely around the edge of our grid. This implies that material that crosses the grid outward will mostly lose its thermal energy to radiation because it will suffer only small adiabatic loses before it crosses the photosphere. 
   
Following this discussion we crudely estimate the light curve to be similar to the rate at which the unbound ejecta carries thermal energy and recombination energy out of the grid. (As we point out below, this does not hold at early times.) We include only the recombination energy of hydrogen in a solar-abundance ejecta as helium recombines deeper inside the ejecta and that energy suffers more adiabatic losses. Overall, we take the contribution of the recombination energy to radiation to be $10^{13} \erg \g^{-1}$. Further adiabatic loses beyond the edge of the numerical grid reduce the contribution of the thermal energy to radiation. On the other hand, parcels of ejecta gas with different velocities (see section \ref{subsec:FlowStructure}) will collide with each other beyond our grid and transfer some kinetic energy to thermal energy and then radiation.    
   
At early times when the NS launches the jets as it enters the envelope we expect to have a bright peak, similar to a shock break-out, because of the initially low-density ejecta. 
However, the exact light curve depends on how the jets' power rises from zero to the value we assume at the beginning. 
We expect that the rise to maximum luminosity will take about a month (rather than the sharp rise in the thermal energy outflow at the edge of the numerical grid that we study next), 
and might contain a peak on this rise. We cannot deal with the first one month (high energy simulations) to a few months (lower energy simulations) of the light curve with our present tools.  

Another potential observational diagnostic might come from X-ray emission from the mass-accreting NS just before it enters the envelope. At this phase the NS accretes from the envelope just before it enters the envelope and before the jets eject envelope gas that will absorb all X-ray emission. Observer on the side where the NS enters the envelope might detect X-ray emission for a fraction of the time the NS crosses the envelope. We cannot estimate the exact time and the X-ray luminosity, but we expect the duration to be several days to few weeks and the X-ray emission to be lower that the visible luminosity during the event.  

We stop the simulation analysis at $t=10^8 \s$ for two reasons. Firstly, the energy outflow rates become very small by that time, below the uncertainties of our simulations. Secondly, the simulations become less accurate at late times because we do not evolve the stellar remnant self-consistently. We expect that the highly disturbed stellar remnant will have a very strong radiation that might blow a strong wind. Nonetheless, we did continue the simulations J41/G1, J42/G2 and J43/G2 for later times to estimate the amounts of mass lost till $t=2\times 10^8$ in these cases. We give the value sinside square brackets in Table \ref{Table:cases}.

\begin{figure} [ht]
\centering
\includegraphics[width=0.48\textwidth]{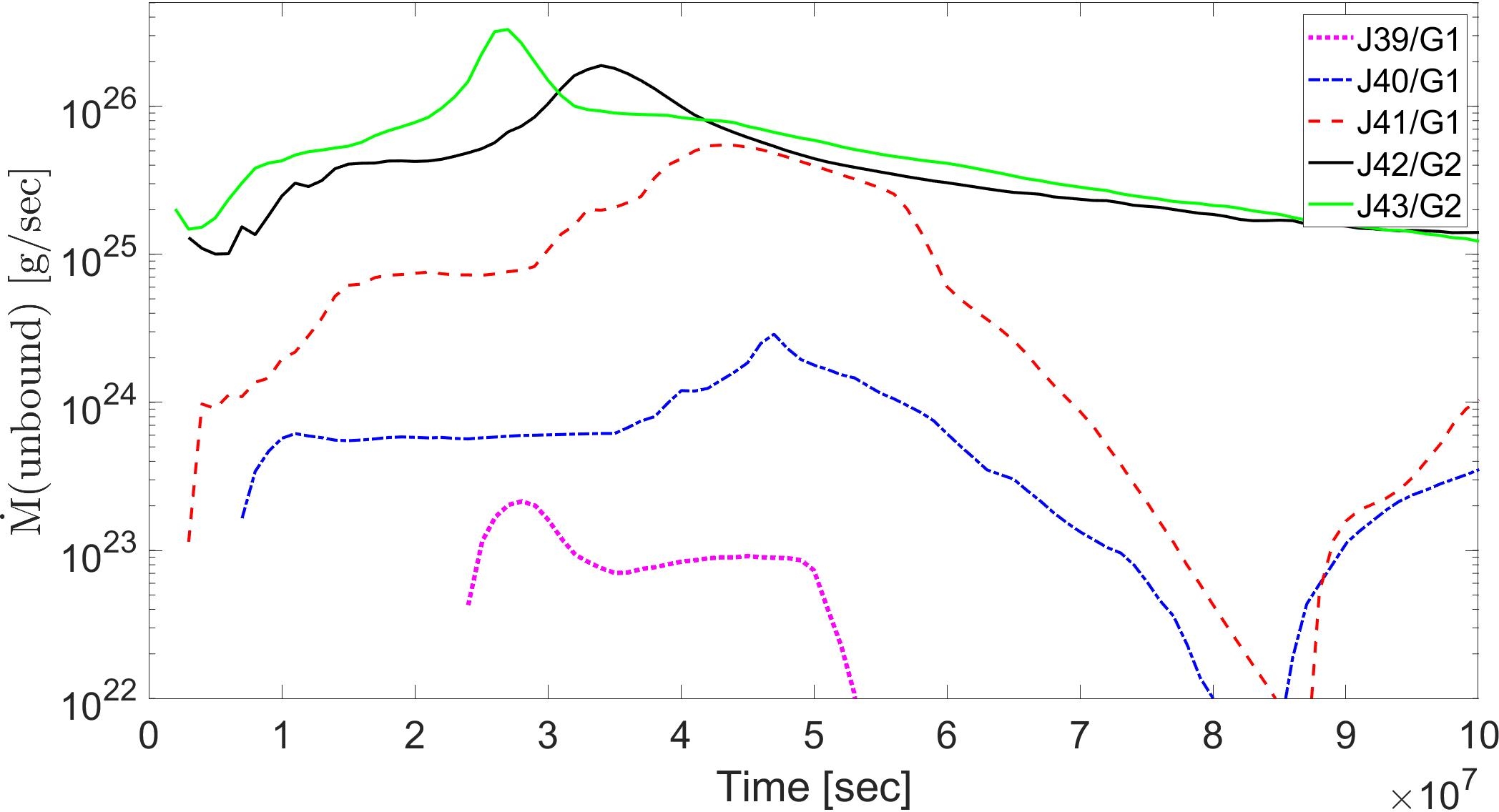}
\includegraphics[width=0.48\textwidth]{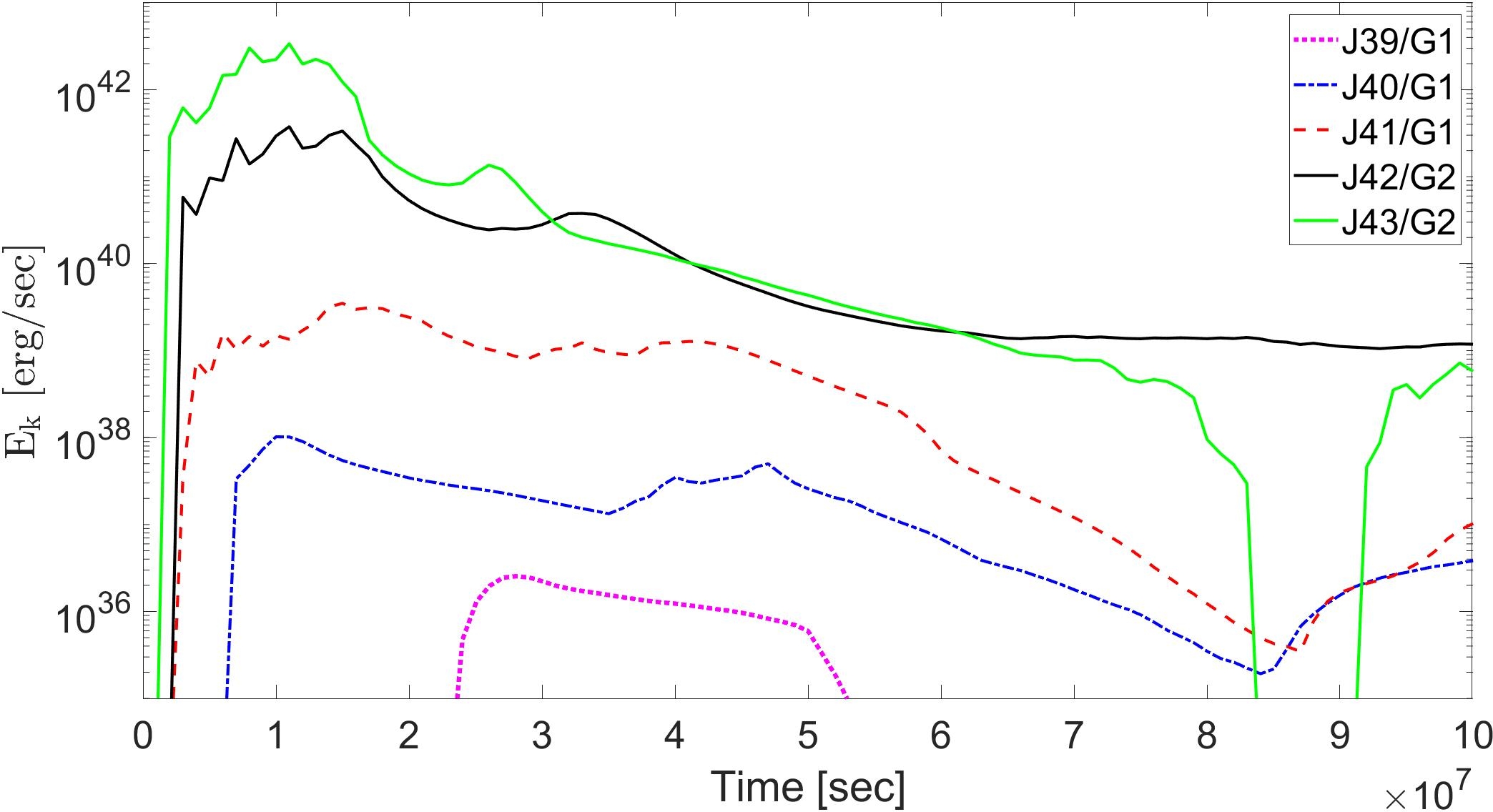}
\includegraphics[width=0.48\textwidth]{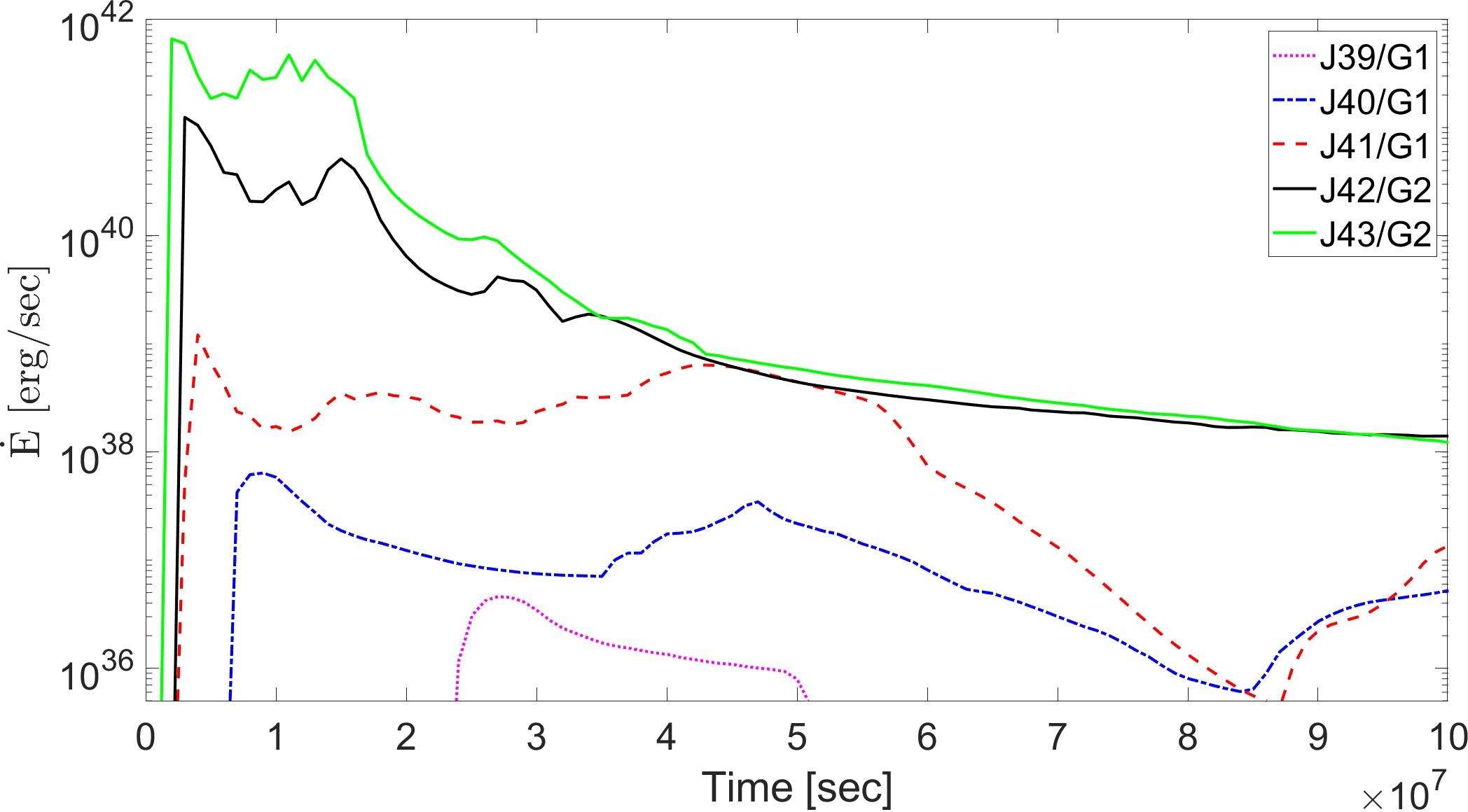}

\caption{The mass outflow rate (upper panel), the kinetic energy outflow rate (second panel), and the outflow rate of the thermal energy plus the hydrogen recombination energy of the outflowing mass (lower panel) for the five simulations. 
  }
\label{fig:OutflowRate}
\end{figure}

In Fig.\ref{fig:OutflowRate} we present the unbound mass outflow rates from the grid (upper panel; the vast majority of the outflowing mass from the grid is unbound), the kinetic energy of the unbound mass outflow rates (middle), and the thermal energy (radiation + gas) + hydrogen recombination energy outflow rates of the unbound mass (lower panel), for the five simulations. We calculate the outflow rates through a sphere of radius $R_{\rm out}=2.5 \times 10^{14} \cm$ in the small grid (the `G1' simulations) and through a sphere of radius $R_{\rm out}=5 \times 10^{14} \cm$ in the large grid (the `G2' simulations).

From the outflow rates that we present in Fig. \ref{fig:OutflowRate} we learn the following. Mass and energy outflow rates are not monotonic but rather suffer ups and downs, i.e., several peaks. The non-negligible mass outflow rate lasts for years (beside the lowest energy simulation J39/G1 where it lasts for about a year), much longer than the jets-activity phase (only about half a year), but shorter than the orbital period we assume in our simulation, $T_{\rm orb}=16.6 \yr$. In any case, the energy that the NS deposits to the envelope and the mass it removes modify the RSG envelope structure that the NS will enter in its subsequent periastron passage. To follow the RSG evolution to the next periastron passage of the NS requires a stellar evolution code, and it is the subject of a future study.  
 
The thermal + recombination energy outflow rates have the following prominent properties that we expect also to be the general properties of the light curves of such transient events. (1) A rise to maximum during one to few months. (2) The luminosity of the first peak (which is the most luminous) is $\approx 10 \%$ of the jets' power at periastron for the energetic simulations, declining to $\la 1\%$ in the least energetic simulations J39/G1, J40/G1, and J41/G1. (3) The light curves slowly decline non-monotonically over time periods of several years (beside in simulation J39/G1 for which it is only about a year).
(4) There are several peaks in the light curves, i.e., `bumpy' light curves. (5) Thermal energy dominates the first part of the light curves and the recombination energy dominates at later times. (6) The total radiated energy (by our assumptions) is $E_{\rm rad}/E_{\rm 2j} \approx 0.3 \%$ of the total energy that the jets carry over their half a year activity period in the lowest power simulation J39/G1, and raises to $E_{\rm rad}/E_{\rm 2j} \approx 4.3 \%$ in simulation J42/G2 (Table \ref{Table:cases}). 

As we explained in section \ref{subsec:Cases} we consider simulation J42/G2 to represent our expectation of the jets' power.  

We here simulated only one NS orbit. Had the NS orbit be further out in the envelope the duration of the event would not change much, but the total event energy be much lower because the NS would accrete from a much lower density zones. Qualitatively, we expect a similar outflow geometry to the lower energy-cases that we simulate here. We expect that a much deeper passage of the NS will result in a much more energetic event that will resemble a CCSN in ejecting large envelope mass. In addition, we expect the energetic jets to largely inflate the remaining bound envelope mass. The large dynamical friction  (gravitational drag) will cause the NS orbit to shrink, such that it might even stay inside the inflated envelope. This is a much more complicated simulation as it requires to include the self-gravity of the envelope.  

\section{Summary}
\label{sec:Summary}

We conducted 3D hydrodynamical simulations of an eccentric CEJSN impostor, i.e., a NS that enters and exits a RSG envelope on a highly eccentric orbit and launches jets as it accretes mass from the RSG envelope. Because the NS does not enter the core and destroy it, this is a CEJSN impostor (rather than a CEJSN). Very similar considerations to those that we presented throughout the paper hold for a BH that enters and exits the RSG envelope, but the mass accretion rate and jets' power are larger. \cite{Gilkisetal2019} present an analytical study of some other properties of CEJSN impostors under different conditions and under some different assumptions.   

We do not have the numerical power to resolve the accretion process onto the NS nor the launching of the jets by the NS. We apply a simple prescription where we inject the assumed jets' power into two opposite conical regions, one on each side of the NS with respect to the orbital plane (section \ref{subsubsec:interaction}). We simulated 5 cases which differ by the combined power of the two jets $\dot E_{\rm 2j}$ (Table \ref{Table:cases}).  For reasons that we discuss in section \ref{subsec:Cases}, we consider simulation J42/G2 to represent the best our expectation of the jets’ power.

We found that the outflow morphology is very complicated (e.g., Fig. \ref{fig:3D_fiducial}). It has a large-scale symmetry about the equatorial plane, but besides that a very large departure from any other symmetry. The outflow morphology might substantially differ between simulations that differ by their jets' power (compare the upper and lower rows of Fig. \ref{fig:outflow_maps} and the two panels of Fig. \ref{fig:outflow_theta}). 
The outflow itself is clumpy and contains many regions with different velocities (e.g., Figs. \ref{fig:rho_v_XY_fiducial}-\ref{fig:J42_vs_J43}). 
The highly non-spherical structure of the ejecta will lead to highly polarised emission, while the clumpy ejecta will lead to `bumpy' light curve. As well, the exact light curve depends on the observer relative direction. 

We plot the mass and energy outflow rates as function of time in Fig. \ref{fig:OutflowRate}. 
We crudely estimate the light curve of these CEJSN impostors to have properties similar to the outflow (from the numerical grid) rate of the thermal energy + hydrogen recombination energy (lower panel of Fig. \ref{fig:OutflowRate}). Under this assumption the efficiency of converting the jets' energy to radiation increases from $ \simeq 0.3 \%$ in the lowest energy simulation J39/G1 to $\simeq 4.3 \%$ in simulation J42/G2.
The light curves have a relatively fast rise (1-few months), but we cannot even crudely estimate this early part of the light curve. Beyond early times, the light curves decay relatively slowly, are `bumpy' with several ups and downs, and they last for about a year to several years. 
Overall, these events will be classified as `gap' objects (\citealt{Gilkisetal2019}; also termed ILOTs), e.g., having a luminosity between those of classical novae and typical SNe.

The ILOT events that we study here occur in massive stars, i.e., young stellar population, last for months to years, and are very energetic and bright. They  have some common properties with ILOTs of very massive stars, e.g., major luminous blue variables eruptions like the 1837-1857 Great Eruption of $\eta$ Carinae (e.g., \citealt{DavidsonHumphreys1997}). The Great Eruption lasted for about 20 years, had several peaks, and amounted to a total kinetic energy of $\approx 10^{50} \erg$ (e.g., \citealt{Smithetal2003}). It is very likely that the companion in $\eta$ Carinae accreted mass and launched jets that powered the Great Eruption (e.g., \citealt{AkashiKashi2020} and references therein). As well, there was a high velocity outflow, $\simeq 10^4 \km \s^{-1}$, during the great eruption (\citealt{Smithetal2018}) that these jets might account for (\citealt{AkashiKashi2020}). Despite these similarities it is possible that we can distinguish between major luminous blue variables eruption and CEJSN impostors by the early light curve (that we have problems to model here), namely, before the system is enshrouded in massive circumbinary gas that forms dust and obscures the binary system. At the time period of days to weeks before the NS enters the envelope and ejects envelope mass, but the NS already accretes mass from the envelope and possibly launches jets, we might detect X-ray, strong UV radiation, and jets with velocities of $\ga 5 \times 10^4 \km \s^{-1}$. 

We did not estimate the rate of such events, as such a calculation requires a population synthesis study that can handle perturbations by a tertiary star. We can only state that such events are very rare. Nonetheless, existing and upcoming sky surveys, e.g., the Zwicky Transient Facility (ZTF; \citealt{Bellmetal2019}), the Large Synoptic Survey Telescope (LSST; \citealt{Ivezicetal2019}), the All-Sky Automated Survey for Supernovae (ASAS-SN; \citealt{Kochaneketal2017PASP}), and the Southern Hemisphere Variability Survey (LSQ; \citealt{Baltayetal2013}), will detect many such transients. In rare cases we expect the event to be CEJSN impostor. We also note that a CEJSN impostor might repeat itself (e.g., \citealt{Gilkisetal2019}).
We repeat the call of \cite{Gilkisetal2019} to  seriously consider  CEJSN impostors as models for peculiar gap objects (ILOTs). 
  
\section*{Acknowledgments}

We thank an anonymous referee for useful suggestions and comments.  
This research was supported by the Pazy Foundation  
SaS thanks National Science Foundation Award 1814967 for its support. FLASH was developed
largely by the DOE-supported ASC/Alliances Center for Astrophysical
Thermonuclear Flashes at the University of Chicago. In producing the images in this paper we used VisIt which is supported by the Department of Energy with funding from the Advanced Simulation and Computing Program and the Scientific Discovery through Advanced Computing Program. This work also required the use and integration of a Python package for astronomy, yt (http://yt-project.org, \citealt{Turk2011}).

\textbf{Data availability}

The data underlying this article will be shared on reasonable request to the corresponding author. 

\label{lastpage}

\begin{thebibliography}{}


\bibitem[Akashi \& Kashi(2020)]{AkashiKashi2020} Akashi, M. \& Kashi, A.\ 2020, \mnras, 494, 3186. doi:10.1093/mnras/staa1014

\bibitem[Armitage \& Livio(2000)]{ArmitageLivio2000} Armitage, P.~J., \& Livio, M.\ 2000, \apj, 532, 540


\bibitem[Baltay et al.(2013)]{Baltayetal2013}  Baltay, C., Rabinowitz, D., Hadjiyska, E., Walker, E.~S., Nugent, P., Coppi, P., Ellman, N.,  et al.\ 2013, \pasp, 125, 683. doi:10.1086/671198 
 
\bibitem[Bellm et al.(2019)]{Bellmetal2019} Bellm, E.~C., Kulkarni, S.~R., Graham, M.~J., Dekany, R., Smith, R.~M., Riddle, R., Masci, F.~J., et al.\ 2019, \pasp, 131, 018002. doi:10.1088/1538-3873/aaecbe 

\bibitem[Chamandy et al.(2018)]{Chamandyetal2018} Chamandy, L., Frank, A., Blackman, E.~G., et al.\ 2018, \mnras, 480, 1898
 
\bibitem[Chevalier(1993)]{Chevalier1993} Chevalier, R.~A.\ 1993, \apjl, 411, L33

\bibitem[Chevalier(2012)]{Chevalier2012} Chevalier, R.~A.\ 2012, \apjl, 752, L2

\bibitem[Davidson \& Humphreys(1997)]{DavidsonHumphreys1997} Davidson, K. \& Humphreys, R.~M.\ 1997, \araa, 35, 1. doi:10.1146/annurev.astro.35.1.1

\bibitem[Fragos et al.(2019)]{Fragosetal2019} Fragos, T., Andrews, J.~J., Ramirez-Ruiz, E., Meynet, G., Kalogera, V., Taam, R.~E., \& Zezas, A., \ 2019, \apjl, 883, L45. doi:10.3847/2041-8213/ab40d1

\bibitem[Fryer et al.(1996)]{Fryeretal1996} Fryer, C.~L., Benz, W., \& Herant, M.\ 1996, \apj, 460, 801

\bibitem[Fryxell et al.(2000)] {Fryxelletal2000} Fryxell, B., et al. 2000, \apjs, 131, 273

\bibitem[Garc{\'\i}a et al.(2021)]{Garciaetal2021} Garc{\'\i}a, F., Simaz Bunzel, A., Chaty, S., Porter, E., \& Chassande-Mottin, E.\ 2021,  \aap, 649, A114. doi:10.1051/0004-6361/202038357

\bibitem[Gilkis et al.(2019)]{Gilkisetal2019} Gilkis, A., Soker, N., \& Kashi, A.\ 2019, \mnras, 482, 4233

\bibitem[Glanz \& Perets(2021)]{GlanzPerets2021} Glanz, H. \& Perets, H.~B.\ 2021, arXiv:2105.02227
  
\bibitem[Grichener et al.(2021)]{Gricheneretal2021} Grichener, A., Cohen, C., \& Soker, N.\ 2021, arXiv:2107.07856

\bibitem[Grichener \& Soker(2019a)]{GrichenerSoker2019a} Grichener, A., \& Soker, N.\ 2019b, \apj, 878, 24

\bibitem[Grichener \& Soker(2019b)]{GrichenerSoker2019b} Grichener, A. \& Soker, N.\ 2019b, arXiv:1909.06328

\bibitem[Grichener \& Soker(2020)]{GrichenerSoker2020} Grichener, A. \& Soker, N.\ 2020, work presented at the EAS Annual Meeting (EWASS) 2020. 

\bibitem[Grichener \& Soker(2021)]{GrichenerSoker2021} Grichener, A. \& Soker, N.\ 2021, arXiv:2101.05118

\bibitem[Holgado et al.(2021)]{Holgadoetal2021} Holgado, A.~M., Silva, H.~O., Ricker, P.~M., \& Yunes, N., \ 2021, \apjl, 910, L22. doi:10.3847/2041-8213/abecdd

\bibitem[Houck \& Chevalier(1991)]{HouckChevalier1991} Houck, J.~C., \& Chevalier, R.~A.\ 1991, \apj, 376, 234


\bibitem[Ivezi{\'c} et al.(2019)]{Ivezicetal2019} Ivezi{\'c}, {\v{Z}}., Kahn, S.~M., Tyson, J.~A., Abel, B., Acosta, E., Allsman, R., Alonso, D., et al.\ 2019, \apj, 873, 111. doi:10.3847/1538-4357/ab042c 

\bibitem[Kochanek et al.(2017)]{Kochaneketal2017PASP}  Kochanek, C.~S., Shappee, B.~J., Stanek, K.~Z.,  Holoien, T.~W.-S., Thompson, T.~A., Prieto, J.~L., Dong S., et al.\ 2017, \pasp, 129, 104502. doi:10.1088/1538-3873/aa80d9

\bibitem[Livio et al.(1986)]{Livioetal1986} Livio, M., Soker, N., de Kool, M., \& Savonije, G.~J., \ 1986, \mnras, 222, 235. doi:10.1093/mnras/222.2.235    
    
\bibitem[Lombardi et al.(2006)]{Lombardietal2006} Lombardi, J.~C., Jr., Proulx, Z.~F., Dooley, K.~L., Theriault, E.~M., Ivanova, N., \& Rasio, F.~A.\ 2006, \apj, 640, 441

\bibitem[L{\'o}pez-C{\'a}mara et al.(2019)]{LopezCamaraetal2019} L{\'o}pez-C{\'a}mara, D., De Colle, F., \& Moreno M{\'e}ndez, E.\ 2019, \mnras, 482, 3646

\bibitem[L{\'o}pez-C{\'a}mara et al.(2020)]{LopezCamaraetal2020MN} L{\'o}pez-C{\'a}mara, D., Moreno M{\'e}ndez, E., \& De Colle, F.\ 2020, \mnras, 497, 2057

\bibitem[MacLeod \& Ramirez-Ruiz(2015a)]{MacLeodRamirezRuiz2015a} MacLeod, M., \&  {Ramirez-Ruiz}, E.\ 2015a, \apjl, 798, L19

\bibitem[MacLeod et al.(2017)]{MacLeodetal2017} MacLeod, M., Antoni, A., {Murguia-Berthier}, A., Macias, P., \&  {Ramirez-Ruiz}, E.\ 2017, \apj, 838, 56

\bibitem[MacLeod \& Ramirez-Ruiz(2015b)]{MacLeodRamirezRuiz2015b} MacLeod, M., \& Ramirez-Ruiz, E.\ 2015b, \apj, 803, 41

\bibitem[Moreno M{\'e}ndez et al.(2017)]{MorenoMendezetal2017} Moreno M{\'e}ndez, E., L{\'o}pez-C{\'a}mara, D., \& De Colle, F.\ 2017, \mnras, 470, 2929

\bibitem[Papish et al.(2015)]{Papishetal2015} Papish, O., Soker, N., \& Bukay, I.\ 2015, \mnras, 449, 288

\bibitem[Paxton et al.(2011)]{Paxtonetal2011} Paxton, B., Bildsten, L., Dotter, A., et al.\ 2011, \apjs, 192, 3

\bibitem[Paxton et al.(2013)] {Paxtonetal2013} Paxton, B., Cantiello, M., Arras, P., et al. 2013, \apjs, 208, 4

\bibitem[Paxton et al.(2015)]{Paxtonetal2015} Paxton, B., Marchant, P., Schwab, J., et al.\ 2015, \apjs, 220, 15

\bibitem[Paxton et al.(2018)]{Paxtonetal2018} Paxton, B., Schwab, J., Bauer, E.~B., et al.\ 2018, \apjs, 234, 34

\bibitem[Paxton et al.(2019)]{Paxtonetal2019} Paxton, B., Smolec, R., Schwab, J., et al.\ 2019, \apjs, 243, 10, 

\bibitem[Rasio \& Shapiro(1991)]{RasioShapiro1991} Rasio, F.~A., \& Shapiro, S.~L.\ 1991, \apj, 377, 559

\bibitem[Ricker \& Taam(2008)]{RickerTaam2008}  Ricker, P.~M., \& Taam, R.~E.\ 2008, \apjl, 672, L41

\bibitem[Schr{\o}der et al.(2020)]{Schroderetal2020} Schr{\o}der, S.~L., MacLeod, M., Loeb, A., et al.\ 2020, \apj, 892, 13

\bibitem[Shiber et al.(2019)]{Shiberetal2019} Shiber, S., Iaconi, R., De Marco, O., \& Soker, N.\ 2019, \mnras, 488, 5615. doi:10.1093/mnras/stz2013

\bibitem[Shiber et al.(2016)]{Shiberetal2016} Shiber, S., Schreier, R., \& Soker, N.\ 2016, RAA, 16, 117

\bibitem[Smith et al.(2003)]{Smithetal2003} Smith, N., Gehrz, R.~D., Hinz, P.~M., Hoffmann, W.~F., Hora, J.~L., Mamajek, E.~E., \& Meyer, M.~R.\ 2003, \aj, 125, 1458. doi:10.1086/346278

\bibitem[Smith et al.(2018)]{Smithetal2018} Smith, N., Rest, A., Andrews, J.~E., et al.\ 2018, \mnras, 480, 1457. doi:10.1093/mnras/sty1479

\bibitem[Soker(2016)]{Soker2016Rev} Soker, N.\ 2016, \nar, 75, 1. doi:10.1016/j.newar.2016.08.002


\bibitem[Soker(2021a)]{Soker2021effervescent} Soker, N.\ 2021a, \apj, 906, 1. doi:10.3847/1538-4357/abca8f

\bibitem[Soker(2021b)]{Soker2021Triple}  Soker, N.\ 2021b, \mnras. doi:10.1093/mnras/stab1275

\bibitem[Soker(2021c)]{Soker2021NSNS} Soker, N.\ 2021c, arXiv:2105.06452

\bibitem[Soker \& Gilkis(2018)]{SokerGilkis2018} Soker, N., \& Gilkis, A.\ 2018, \mnras, 475, 1198

\bibitem[Soker et al.(2019)]{Sokeretal2019CEJSN} Soker, N., Grichener, A., \& Gilkis, A.\ 2019, \mnras, 484, 4972

\bibitem[Turk et al.(2011)]{Turk2011} Turk, M.~J., Smith, B.~D., Oishi, J.~S., et al.\ 2011, \apjs, 192, 9 

\bibitem[Vick et al.(2021)]{Vicketal2021MNRAS} Vick, M., MacLeod, M., Lai, D., \& Loeb A.\ 2021, \mnras, 503, 5569. doi:10.1093/mnras/stab850

\bibitem[Zevin et al.(2021)]{Zevinetal2021} Zevin, M., Bavera, S.~S., Berry, C.~P.~L., et al.\ 2021, \apj, 910, 152. doi:10.3847/1538-4357/abe40e

\bibitem[Zheng \& Yu(2021)]{ZhengYu2021} Zheng, J.-H. \& Yu, Y.-W.\ 2021, arXiv:2103.15576

\end{thebibliography}
\end{document}